\documentclass[iop]{emulateapj}
\usepackage{lscape}
\usepackage{rotating}
  \received{}
  \revised{}
  \accepted{}
  
\def\lsim{\mathrel{\rlap{\lower3.5pt\hbox{\hskip0.5pt$\sim$}}
    \raise0.5pt\hbox{$<$}}}                
\def\gsim{~\rlap{$>$}{\lower 1.0ex\hbox{$\sim$}}}
\shortauthors{Romeo et al.} \shorttitle{Mass-metallicity evolution}

\begin{document}

\title{Evolution of the Mass-Metallicity relations in passive and star-forming galaxies from SPH-cosmological simulations }

\author{A.D. Romeo Velon\`a\altaffilmark{1}, J. Sommer-Larsen\altaffilmark{5,6,7}, N.R. Napolitano\altaffilmark{2}, 
V. Antonuccio-Delogu\altaffilmark{3}, 
S. Cielo \altaffilmark{4}, I. Gavignaud\altaffilmark{1}}

\altaffiltext{1}{
Universidad Andres Bello, Departamento de Ciencias F\'isicas, Av. Rep\'ublica 220, Santiago, Chile}\email{\texttt
aro@oact.inaf.it}
\altaffiltext{2}{
INAF -- Osservatorio Astronomico di Capodimonte, Salita Moiariello 16, I-80131, Napoli, Italy}
\altaffiltext{3}{
INAF -- Osservatorio Astrofisico di Catania, v. S.Sofia 78, I-95123, Catania, Italy}
\altaffiltext{4}{
Max-Planck-Institut f$\rm \ddot{u}$r Astronomie K$\rm \ddot{o}$nigstuhl 17, D-69117 Heidelberg, Germany}
\altaffiltext{5}{
Dark Cosmology Centre, Niels Bohr Institute, University of Copenhagen, Juliane Mariesvej 30, DK-2100 Copenhagen, Denmark}
\altaffiltext{6}
{Excellence Cluster Universe, Technische Universit\"at M\"unchen, Boltzmannstrasse 2, 85748 Garching, Germany}
\altaffiltext{7}
{Marie Kruses Skole, Stavnsholtvej 29-31, DK-3520 Farum, Denmark}

\begin{abstract}
We present results from SPH-cosmological simulations, including self-consistent modelling of SN feedback and chemical evolution, 
of galaxies belonging to two clusters and twelve groups. 
We reproduce the mass--metallicity (ZM) relation of galaxies classified in two samples according to their star-forming activity,
as parametrized by their sSFR, across a redshift range up to $z$=2.

The overall ZM relation for the composite population evolves according to a redshift-dependent quadratic functional form
that is consistent with other empirical estimates, provided that the highest mass bin of the BCGs is excluded.

Its slope shows irrelevant evolution in the passive sample, being steeper in groups than in clusters.
However, the sub-sample of high-mass passive galaxies only is characterized by a steep increase of the slope with redshift,
from which it can be inferred that the bulk of the slope evolution of the ZM relation is driven by the more massive passive objects.
The scatter of the passive sample is dominated by low-mass galaxies at all redshifts and keeps constant over cosmic times.
The mean metallicity is highest in cluster cores and lowest in normal groups, following the same environmental sequence 
as that previously found in the Red Sequence building.

The ZM relation for the star-forming sample reveals an increasing scatter with redshift, indicating that it is still being built at early epochs.
The star-forming galaxies make up a tight sequence in the SFR-$M_*$ plane at high redshift, whose scatter increases with time alongside with the consolidation of the passive sequence. We also confirm the anti-correlation between sSFR and stellar mass, pointing at a key role of the former in determining the galaxy downsizing, as the most significant means of diagnostics of the star formation efficiency. Likewise, an anti-correlation between sSFR and metallicity can be established
for the star-forming galaxies, 
while on the contrary more active galaxies in terms of simple SFR are also metal-richer.

Finally the [O/Fe] abundance ratio is presented too: we report a strong increasing evolution with redshift at given mass, especially
at $z\gsim$1. The expected increasing trend with mass is recovered when only considering the more massive galaxies.

We discuss these results in terms of the mechanisms driving the evolution within the high- and low-mass regimes
at different epochs: mergers, feedback-driven outflows and the intrinsic variation of the star formation efficiency.
\end{abstract}


\section{Introduction}

Nearby galaxy clusters are dominated by bright, massive early-type (ET) galaxies, which mostly consist of old stellar populations.
Models of galaxy formation are strongly constrained by star formation histories of ET galaxies, that in turn have been
probed by different observational means of diagnostics. One traditional test for galaxy formation theories 
is the colour-magnitude relation (CMr) -and more specifically the building-up of the so-called Red Sequence (RS) of ET galaxies. 
Clusters ellipticals form a tight RS, which is classically interpreted as a mass--metallicity (hereafter ZM) relation 
(see Bower, Lucey \& Ellis 1992, Gladders et al. 1998, Hogg et al. 2004, McIntosh et al. 2005).

A complete model of galaxy formation has to face up the question of reproducing both the photometric and spectroscopic observables 
--for example Pipino \& Matteucci (2008) have shown that the bulk of star formation and the galaxy assembly should 
occur simultaneously in order to reproduce at the same time all the chemical properties of present-day massive 
ellipticals, 
like the observed relations of [Fe/H]--mass and [Mg/Fe]--mass (as in Thomas et al. 2005). 
The latter relation stems from a collateral manifestation of the {\it downsizing} (so called chemo-archaeological), in which more massive 
galaxies present higher $[\alpha/Fe]$ ratios (Thomas 1999) --a trend interpreted with shorter formation timescales in more luminous 
objects (Matteucci 1994).

More widespread, the local ZM relation holds for all galaxy types, with the more massive being the metal richer.
Several observational studies have extended the ZM relation up to $z$=1, finding a flattening with redshift
(Savaglio et al. 2005 -hereafter S05, Lamareille et al. 2009)
and also with mass at any given $z$, in the sense that the relation forms a plateau at high masses and gets steeper
at the faint end. 
Erb et al. (2006, hereafter Erb06), Maiolino et al. (2008, hereafter M08), Mannucci et al. (2010) have found a strong and monotonic 
evolution of the ZM relation, with metallicity decreasing with redshift at a given mass and interpreting this as a side-effect of a
a metallicity--gas fraction relation holding at $z>$2.
In particular when going to even higher redshifts ($z\gsim$3), M08 found evidences for a stronger
evolution at all masses, but especially for less massive galaxies -what they attribute to {\it downsizing} effect.
Likewise, semi-analytical works such as Vale Asari et al. (2009, hereafetr VA09) found that more massive galaxies show
very little evolution since a lookback time of 9 $Gyr$, having evolved fastly in the past -which supports again the {\it downsizing} framework.
On the other side, P\'erez-Montero et al. (hereafter PM09) found a flattening at $z$=0.9-1.2 with respect to the local relation,
with most of the evolution driven by most massive galaxies: they ascribe this as a result of lower effective yields in more massive
galaxies at higher $z$, accordingly to the hierarchical scenario.

Within the hierarchical model, hydrodynamical simulations with chemical enrichment (but without SN winds) by Tissera, De Rossi \& Scannapieco 
(2005, hereafter TDRS05) and De Rossi, Tissera \& Scannapieco (2007) confirmed a ZM relation well
established already at $z\sim$3 and weakly evolving thereafter; its slope changes at a characteristic galaxy mass
of $3\times 10^{10}M_{\odot}$, below which the evolution is ruled by wet mergers affecting the resulting metallicity.
Other cosmological simulations, including winds, by Dav\'e, Finlator \& Oppenheimer (2011) explained the slow dropping of gas fractions and slow
rise of metallicity with time (at given mass) as a result of an increasing metallicity in the accreted gas, inflowing at a rate 
that decreases faster than the gas depletion rate.

In general the origin of a ZM sequence has been linked to the higher efficiency of SN outflows at expelling
metal-enriched gas in lower mass galaxies (see Arimoto \& Yoshi 1987, Gallazzi et al. 2005, Kobayashi, Springel 
\& White 2007). On the other side, the connection between star formation efficiency and ZM relation was evidenced, for example, by 
Tremonti et al. (2004), who found that star-forming galaxies at $z$=0.1 exhibit a strong ZM relation.
Brooks et al. (2007) and Finlator \& Dav\'e (2008) also concluded that SF efficiency is the primary driver of the ZM relation.
In particular, Calura et al. (2009) demonstrated that it can arise as a natural by-product of the increasing 
efficiency of star formation (parametrized as SFR per unit mass of gas) with galaxy mass, without need to invoke
gas outflows producing metal losses in less massive galaxies. Similar conclusion were reached by VA09, whilst Gallazzi et al. (2005) 
argued that neglecting winds does not correctly reproduce the observations.

In any case, toy models without SN feedback nor winds (so called {\it closed box}) have proven to result into early metal overproduction
in low-mass galaxies and hence a too flat ZM relation (see Brooks et al. 2007), or into a general overestimated metallicity
at all masses at high redshift (De Rossi, Tissera \& Scannapieco 2007, Finlator \& Dav\'e 2008).
After all, outflows are to be considered as a natural and necessary mechanism to expel metals from the galaxy, yet
their effects on the slope of the ZM relation are still not clear.
In general the latter depends on the effective metal yield (in turn depending on the cold gas fraction
available to star formation) and the combination of inflows and outflows: under this respect, S05 found that a closed box model
results into a steeper relation, and Erb06 concluded as well that their $z\sim$2 data are better described by a shallower slope
as the one deriving from significant outflows.

Ellison et al. (2008: hereafter E08) and Mannucci et al. (2010) have pointed at a relation between
specific SFR (sSFR) and metallicity as a more general scaling relation in a threefold space including mass, SFR and metallicity.
Romeo et al. (2008, hereafter R08) have extended this dependency to the colour-magnitude relation itself, which the ZM relation can be 
considered as an undegenerate projection of: therein the sSFR was proposed as a third variable to describe its evolution.
P\'erez-Montero et al. (2013, hereafter PM13) recently correct the ZM relation to take into account the natural evolution of the SFR with redshift,
consistently using for the first time a composite sample of galaxies spanning over a redshift range up to $z$=1.3.
 
Indeed discrepancies arise in the available data about the dependence of the ZM relation on the SFR:
relatively to local SDSS samples, E08 had deduced a dependence of the ZM relation on sSFR at lower masses 
($M_*<10^{10}M_\odot$), according to which galaxies with higher sSFR
have lower gas-phase metallicities for a given stellar mass by a factor $\sim$0.1 dex respect to the less star-forming; they
impute this to different star formation efficiencies, in the sense that more intense past star-forming activity would yield higher present gas metallicity
but lower current sSFR, inasmuch as more gas had been previously depleted. 
In sharp contrast to them, Lara-L\'opez et al. (2010) found instead a shallow yet positive correlation between SFR and metallicity at low redshift, 
by virtue of which high SFR galaxies tend to have higher metallicities. 
Yates, Kauffmann \& Guo (2012) explain this result with a semi-analytical model by which passive and massive galaxies have exhausted
their gas reservoirs during a recent major merger, that inhibited further star formation.

\bigskip

All in all, there remain still open two compelling questions about the evolution of the ZM relation: 1) whether it is mainly driven by
galactic winds or just by the natural trend of the galaxy star formation history, and 2) whether the variation of its slope with redshift
is more compatible with {\it downsizing} or with hierarchical assembling of stellar mass in galaxies.
In this paper we aim at studying both the absolute variation of metallicity with mass at different redshifts (the {\it
evolution} of the ZM relation), and the evolution of its slope as well. The latter gives more precise indications about how the
almost monotonic increase of metallicity with time at given mass depends in turn on the galaxy mass itself, hence represents
an optimal diagnostic tool to shed light on both the questions aforementioned.

In this work we will consider two galaxy classes, according to their sSFR: star-forming and passive galaxies, separated by redshift-dependent
thresholds mimicking the observed RS evolution (see Section 3). 
Caution has to be warned about coping with passive and star-forming objects within the same framework, since metallicity is measured by
different methods in either of these: 
most of the observational data we are going to compare with, refer to inter-stellar galactic metallicity, that is commonly expressed in terms of oxygen abundance and derived 
from nebular emission lines in gas-rich or star-forming galaxies. For local ellipticals instead a luminosity-weighted stellar ZM is rather employed 
by aid of SSP synthesis models (see Thomas et al. 2007).
Current semi-analytical studies can employ either the stellar metallicity $Z_*$ (e.g. Pipino et al. 2009 and VA09) or the gaseous one (Sakstein et al. 2011).
In general, besides to being less prone to spectroscopic bias, as an intrinsic parameter the stellar metallicity allows to probe metallicities at
different epochs of galaxy evolution, which is this paper's main scope.

Moreover, Gallazzi et al. (2005) have demonstrated a ZM relation holding with stellar metallicities,
even though with an offset to lower values than nebular ones. Finlator \& Dav\'e (2008) confirmed that
the metallicity of the parent gas cloud is mostly tracked by younger galaxies, whose
UV-only luminosity-weighted metallicity overestimates by 50\% the mean values in SSPs.
The main difference is that bolometric stellar metallicity traces all the past metal enrichment as result of an
extended star formation history, whereas nebular metallicity (or UV-only luminosity-weighted metallicity) 
rather only mirrors the more recent star forming episodes (see also Sommariva et al. 2012). 
As a consequence, we prefer to rely on stellar metallicities as a more suitable
parameter to combine with the other relevant one: the specific star formation rate 
(sSFR), defined as the aggregate SFR over the {\it final} stellar mass of the galaxy at that redshift, 
i.e. calculated at the same time as the galaxy age. 

\begin{figure*}
\centering
\includegraphics[width=15cm,height=10cm]{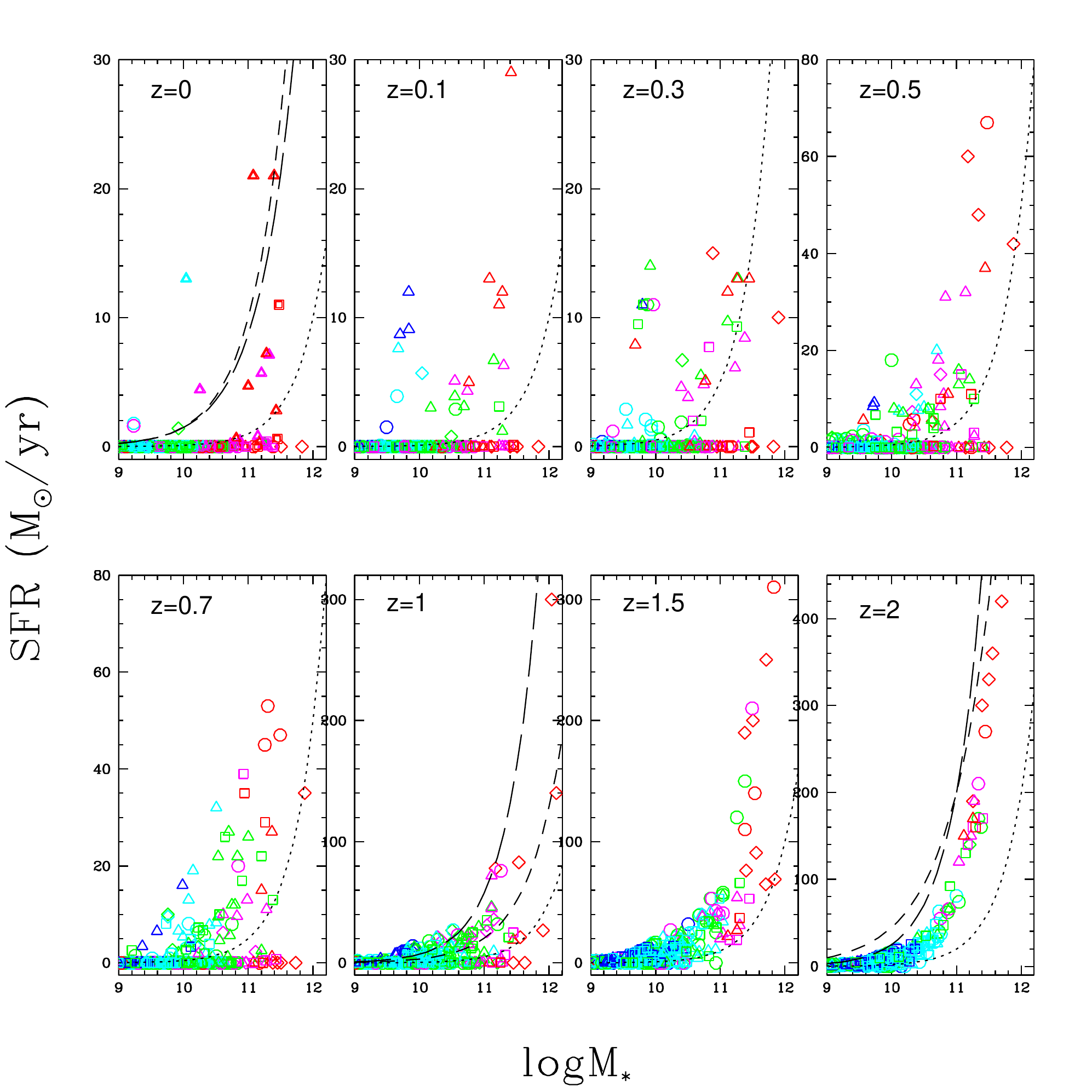}
\caption{Relation between SFR and stellar mass at different redshifts, for the four galaxy classes considered: cluster IN (diamonds), cluster OUT (circles),
normal groups (triangles), fossil groups (squares), all classified by their mass-weighted total metallicity $Z_*/Z_\odot$: blue ($<$0.5), cyan (0.5-1), green (1-1.5),
magenta (1.5-2), red ($>$2).
Data (long-dashed lines) at $z$=0 are from SDSS, at $z$=1 (Elbaz et al. 2007) and $z$=2 (Daddi et al. 2007) from GOODS --all for star-forming galaxies;
all are compared with Millennium simulation (Kitzbichler \& White 2007, dashed).
The dotted curve gives the threshold in sSFR considered to select the active and passive samples (see Section 3). }
\label{SFR}
\end{figure*}

\begin{figure*}
\centering
\includegraphics[width=15cm,height=10cm]{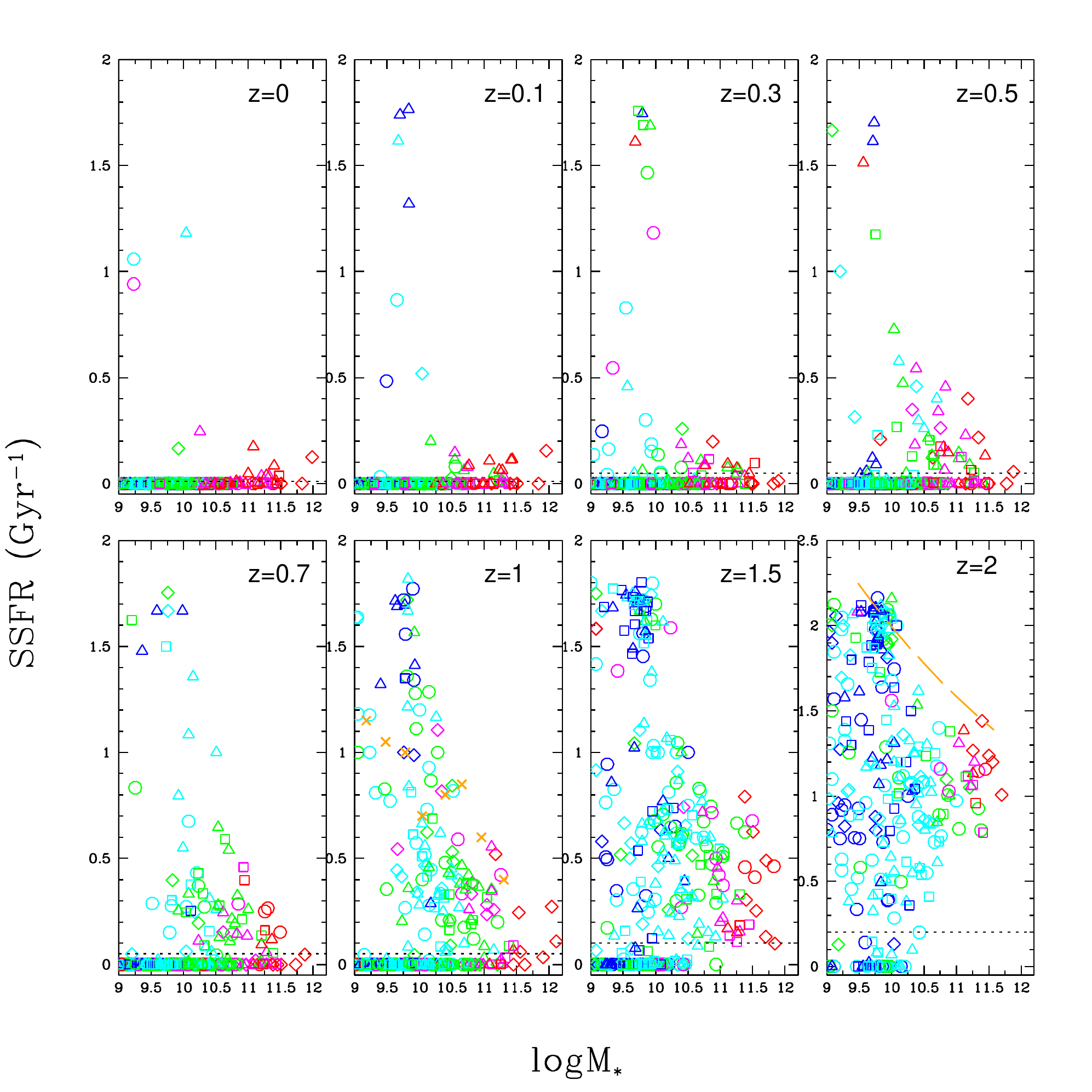}
\caption{Relation between specific SFR (defined as $SFR/M_*$) and stellar mass at different redshifts, for the four galaxy classes considered (same symbols as previous figure).
Data at $z$=1 (orange crosses, Elbaz et al. 2007) and $z$=2 (orange dashed curve, Daddi et al. 2007) are from star-forming galaxies in GOODS. 
The dotted horizontal line gives the threshold in sSFR considered to select the active and passive samples (see Section 3).}
\label{sSFR}
\end{figure*}

\begin{figure*}
\centering
\includegraphics[width=15cm,height=10cm]{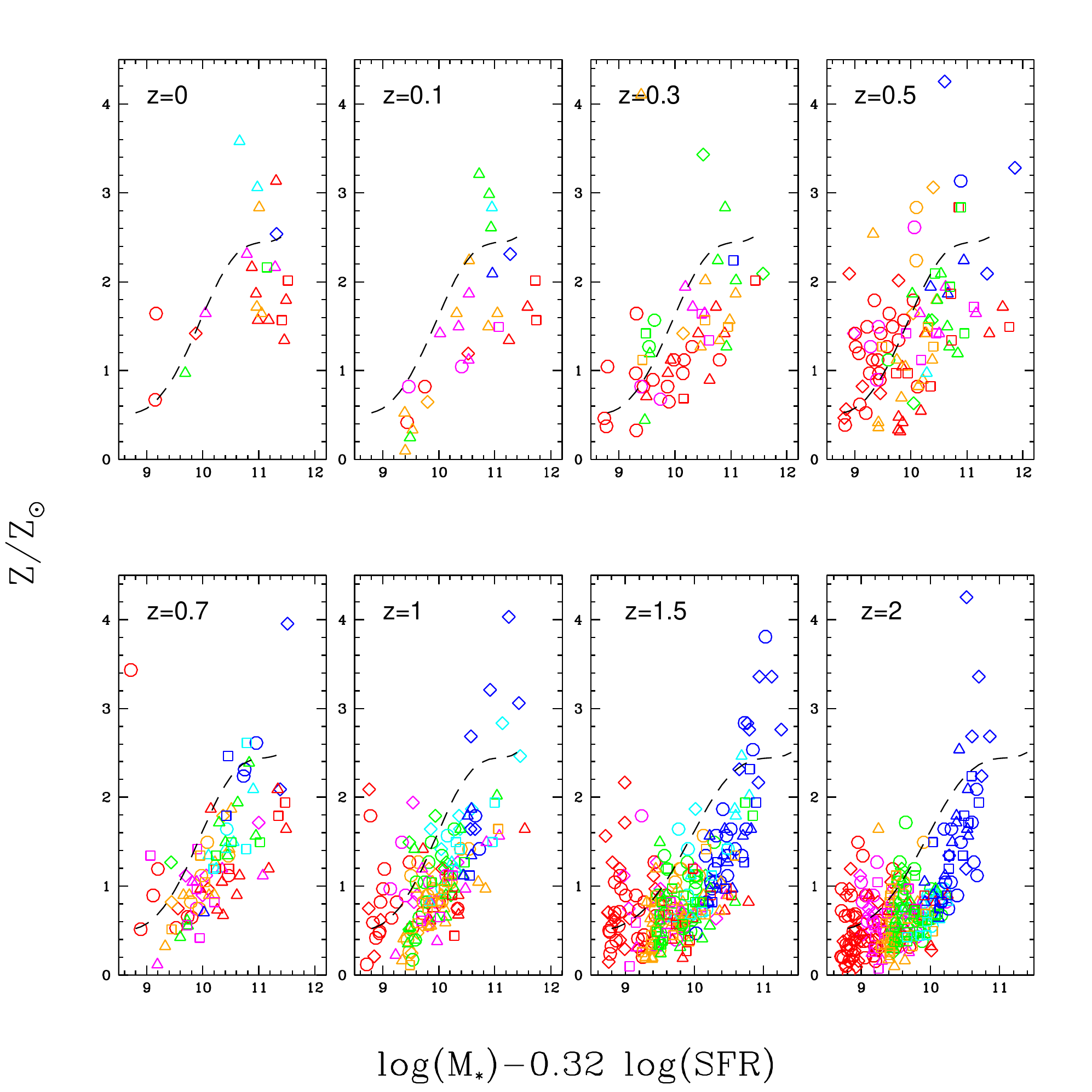}
\caption{Relation between stellar metallicity and the parameter $\mu_{min}=logM_*-0.32log(SFR)$ as defined in Mannucci et al. (2010): the polynomial fit proposed
therein is plotted as dashed curve. Galaxies considered here are only the star-forming ones, belonging to the four environmental classes 
(same symbols as previous figures) and coloured according to their SFR: red (SFR$<$2), magenta (2$<$SFR$<$5), orange (5$<$SFR$<$10), green (10$<$SFR$<$20), 
cyan (20$<$SFR$<$30), blue (SFR$>$30), in units of $M_\odot/yr$.}
\label{mu_SFR}
\end{figure*}

\begin{figure*}
\centering
\includegraphics[width=15cm,height=10cm]{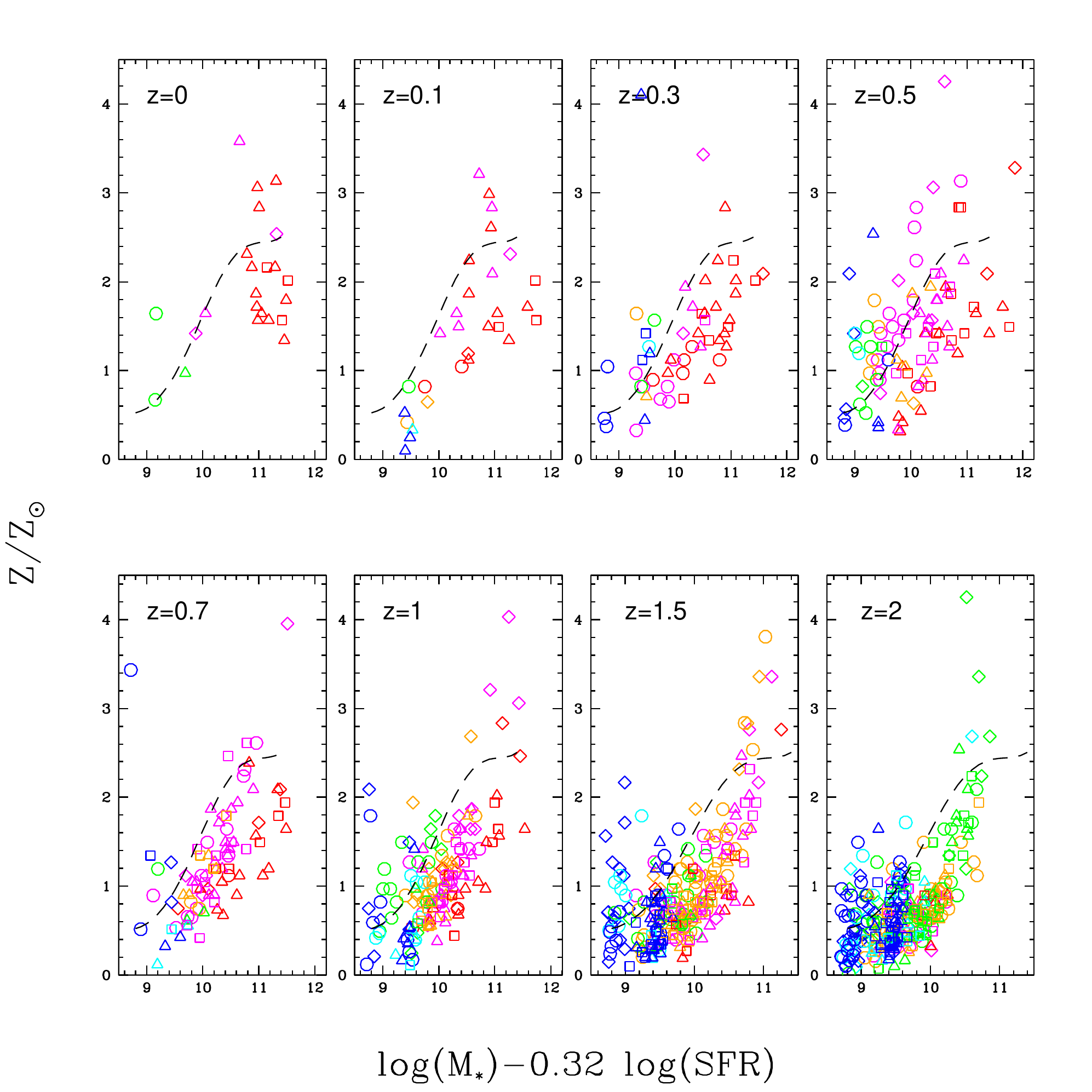}
\caption{Same as previous figure, with galaxies coloured according to their sSFR: red (sSFR$<$0.1), magenta (0.1$<$sSFR$<$0.4), orange (0.4$<$sSFR$<$0.8), green (0.8$<$sSFR$<$1.2), 
cyan (1.2$<$sSFR$<$1.5), blue (sSFR$>$1.5), in units of $Gyr^{-1}$.}
\label{mu_sSFR}
\end{figure*}

\begin{figure*}
\centering
\includegraphics[width=15cm,height=10cm]{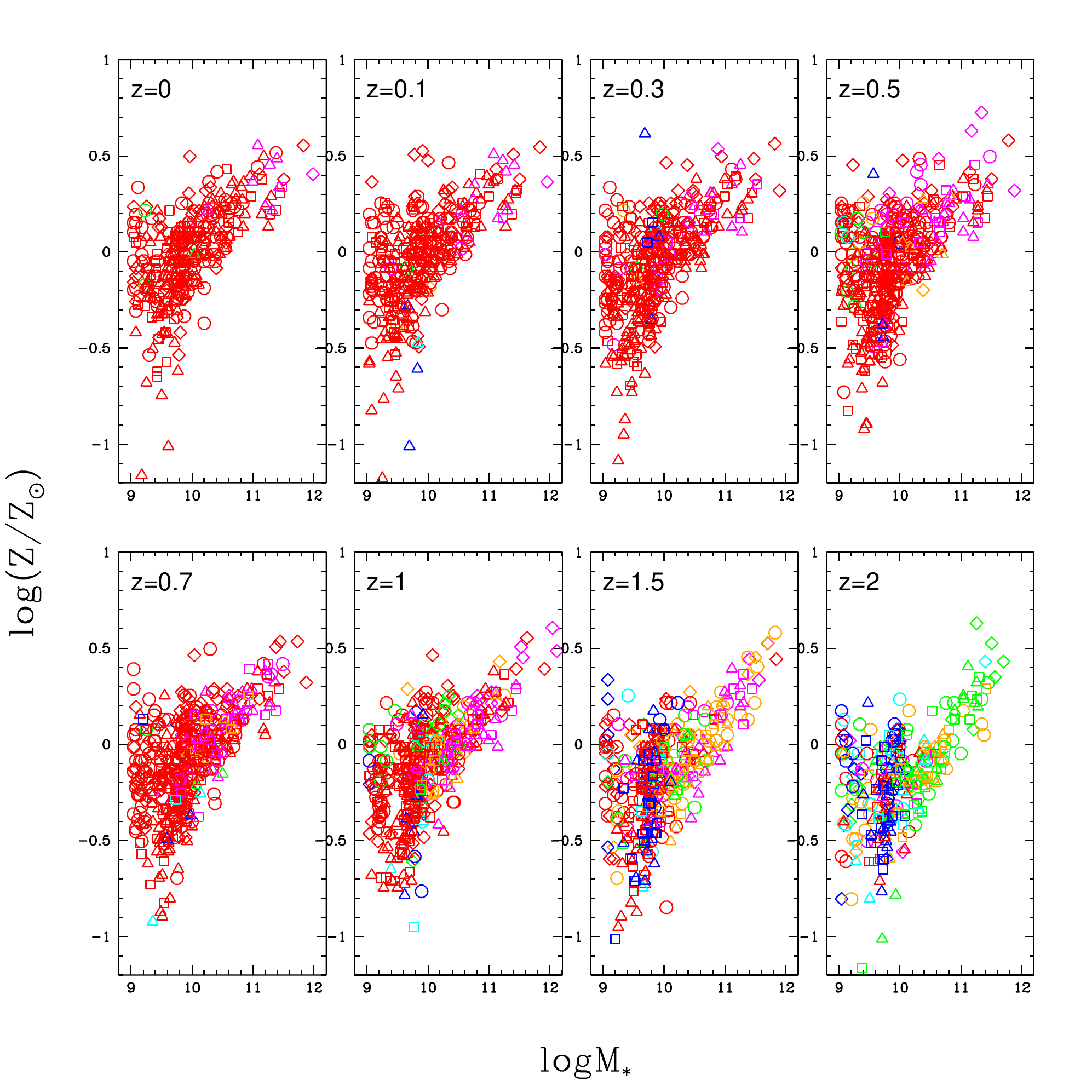}
\caption{ZM relation for all galaxies more massive than $10^9M_\odot$, coloured according to their sSFR: red (DS), magenta (0.1$<$sSFR$<$0.4), orange (0.4$<$sSFR$<$0.8), green (0.8$<$sSFR$<$1.2), 
cyan (1.2$<$sSFR$<$1.5), blue (sSFR$>$1.5), in units of $Gyr^{-1}$.}
\label{ZM_SSFR}
\end{figure*}

\section{Simulations}

In the simulations here presented we consider as standard recipe that with ``super-winds" scheme, 
that has proven to successfully reproduce many of the chemical and thermal properties of the IGM (Romeo et al. 2006:
Figs. 13, 14, 16, 18, 19 therein demonstrate that models with weak stellar feedback are inadequate to account for the abundance 
levels of the IGM in clusters or groups). In particular, the star-to-IGM iron mass ratio in the models without outflows
is strongly skewed towards a five times higher metal content locked up into stars than in the outer gas (at $z$=0); this 
means that whatever {\it closed box} model tends by its nature to heavily overestimate stellar metallicity, as well as the ISM gaseous one
(see below).

Our scheme works as follows: gas and metal galactic outflows are triggered by starburst episodes, their strength being
proportional to the fraction of stars partaking in each starburst.
Each star particle hosts a Single Stellar Population (SSP) of total mass corresponding to the
stellar mass resolution of the simulation, that in this work is $4.43\times 10^7 M_\odot$.
The individual stellar masses are distributed according to an Arimoto-Yoshi IMF.
Each of these SSPs represents a coeval and chemically homogeneous stellar population, characterized by its age and metallicity.
The newborn stars evolve according to the IMF chosen and feed energy back to the ISM mostly by means of SN II:
the energy from SN II is transferred to the ISM as thermal, whence a fraction is converted into kinetic energy of
the expanding fronts.

At the same time, stars return to the surrounding gas also chemically processed material, leading to an increase in its cooling rate.
The resulting stellar feedback is then the combination of energy (given by SN rate) and metal release.
During the SSP lifetime, energy and metals are thus injected into the gas, according to a stochastical return model. 
The whole cycle gas-stars consistently follows
the birth and evolution of the SSPs and their final ``decay'' again into gas particles through feedback
mechanism; the latter will in turn regulate further episodes of star formation, and so forth.
All the chemical yields are calculated within a {\it non-instantaneous recycling} model, taking into account the stellar
timescales.

More details about the cosmological and SPH simulation are available in Romeo et al. 2006 and 2008. 
New galaxy-finding algorithms have been implemented on the top of the original simulations.
It is worth stressing that most of theoretical works on the subject, as well as many observational extrapolations
at higher redshifts, have been developed under significant approximations:
either the {\it closed box} (e.g. VA09, TDRS05, De Rossi, Tissera \& Scannapieco 2007, S05) or
the instantaneous recycling chemical evolution (VA09). 
Remarkable exceptions are for example Dav\'e, Finlator \& Oppenheimer (2011), who ran N-body + SPH simulations
within the same rezooming paradigm as ours and adopting similar physical modules; they focussed however on
field galaxies and in addition experimented variable winds.

The individual galaxies selected are then assigned a luminosity as the sum of the luminosities 
of its constituent star particles, in the classical broad bands {\it UBVRIJHK}.
For each SSP, luminosities are computed by mass-weighted integration of the Padova 
isochrones (Girardi et al. 2002). Thus the physically meaningful quantities of our data
are age and metallicity, from which colours are derived. 
All results are here presented for (total, O and Fe) mass-weighted metallicities, but data are 
also available for Si and Mg and for luminosity-weighted metallicity as well.

\section{Samples and Methods}

The targets selected from the cosmological simulation were one cluster of $T\sim 3$ keV, one of $T\sim 6$ keV
and 12 groups ($T\sim 1.5$ keV), four of which fossil.
The two clusters are in different dynamical stages of their evolution, being the larger less relaxed
and hence dynamically younger (undergoing the last major merger at $z\sim$0.8) than the smaller (for which
the last major merger occurred at $z\sim$1.5); notwithstanding, we have stacked their galaxies together aiming at 
reproducing the natural cosmic variance observed at any epoch.

All simulated galaxies belong to a class of environment, which in the clusters is a function of
radial cluster-centric distance, corresponding to regions inside and outside
a threshold radius of $1/3R_{vir}$ (with $R_{vir}=R_{180}$). All in all, there are four
classes of galaxies: the two cluster cores (IN), the two cluster outskirts (OUT), the eight normal groups
(NG) and the four fossil groups (FG).

The issue of defining a consistent selection of RS members to be compared with observations
at different redshifts, is a crucial one.
To this purpose, we have considered two different galaxy samples:
\\
\indent 1) The {\it dead sequence} sample (DS), which belong to all and only those galaxies having 
sSFR below a threshold that has been defined as slowly progressive with redshift:
from $10^{-2}$ (at $z$=0) to $10^{-1}$ (at $z$=2) $M_\odot Gyr^{-1}/M_*$.
By virtue of its colour-independent definition, the DS may be considered as the model equivalent of the observed RS
(see R08).
\\
\indent 2) The star-forming sample (SF), comprising galaxies having sSFR higher than the above threshold value.
These objects correspond to the {\it blue cloud} in the colour-magnitude plane (see R08).

We are using a separation in specific SFR ($=SFR/M_*$), rather than SFR {\it tout court}, because the former is a parameter closer
to the star formation {\it efficiency}, defined as proportional to $SFR/M_{cold}$ (see e.g. Sakstein et al. 2011).
As such, sSFR allows a more direct analysis of the $SFR-Z-M_*$ plane as introduced by Mannucci et al. (2010).
In fact, if one defines the efficieny such that $SFR=\epsilon M_{gas}$ (from a Schmidt-Kennicutt unit power-law),
then $sSFR=\epsilon \frac{M_{gas}}{M_*}$, that is sSFR and star formation efficiency are linked by a factor which is
function of the gas fraction only (see Lilly et al. 2013).

The redshift dependence of the threshold is empirically mimicking the decreasing SFR with time (see e.g. Lilly et al. 1996),
and is quantitatively equivalent to common parametrizations used in S.A.M. (see again Sakstein et al. 2011).
It is also compatible with observationally established divisions of the CM plane in order to define the RS on a colour
base (see Bell et al. 2004).

\bigskip

As a general matter, any comparison with and between data is affected by both aperture effects and by different metallicity indicators used:
this is true up to the extent that it is seldom possible to place different samples on the same metallicity scale, due to the lack of
conversion factors between the indicators (see Erb06, Foster et al. 2012, PM13).
The estimation of the global trend of the ZM relation at different cosmological epochs is possible only through the analysis
of statistically signiﬁcant samples, provided that consistently adopting the same calibrations: in particular 
PM09 warn against a selection bias towards galaxies with bright emission lines and hence lower metallicities, while
Andrews \& Martini (2013) claim that the commonly used method of strong line ratios needs a complicate empirical
calibration resulting into an indirect approximation to real metallicities.

Due to internal metallicity gradients in galaxies (see Tortora et al., 2011), also the choice of aperture is critical in order to compare with observational results: 
Ellison \& Kewley (2005) calculated that different apertures in the SDSS can account for a $\sim0.15$ dex offset between nuclear and integrated metallicities.
This may lead to an uncertainty in the metallicity which is approximately of a factor twice, that is about the same as the observed difference
between the amplitudes of the ZM relation for $z\sim$2 and local galaxies (Erb06).
However, Ellison et al. (2009, hereafter E09) have demonstrated that the ZM relation with stellar metallicity is much less affected by aperture cuts 
in terms of half light radii, than that with nebular gas metallicity.

Throughout this work, the colours of our brightest central galaxies (BCG, or CD) were corrected using fractional apertures in terms 
of the virial radius (that is, 0.003-0.008$R_{vir}$, being $R_{vir}$ decreasing with redshift: the resulting value is approximately equivalent 
to $R_{eff}$ for the CD), 
in order to cut the diffuse envelope out; thus was excluded the intra-cluster stellar component, whose
profile can often be superposed to the extended BCG one (see Sommer-Larsen, Romeo \& Portinari, 2005, for more details about). 
Moreover we cut the innermost 10 kpc out of the BCGs, since in many cases the presence of excess young stellar populations
therein affects the galaxy colour resulting in a bluer position within the colour-magnitude plane (see R08).
Consequently, the whole CDs were accounted for as to their magnitude and the aperture-corrected ones as to their colour.

Finally, as anticipated in the Introduction, we can choose to model our results in terms of the gaseous or the stellar metallicities, that are 
more suitable to describe, respectively, the star-forming or the passive galaxies.
However, in order to attain a more consistent description that be valid for both samples, and also
due to the larger uncertainties met in binding the gas particles to the galaxy stellar component within a self-consistent radius, 
in the following we will limit our analysis to the stellar metallicities only -with the exception of subsection 4.3, where results
for the gas abundances are presented as well for reference.


\begin{figure}
\includegraphics[width=9cm,clip]{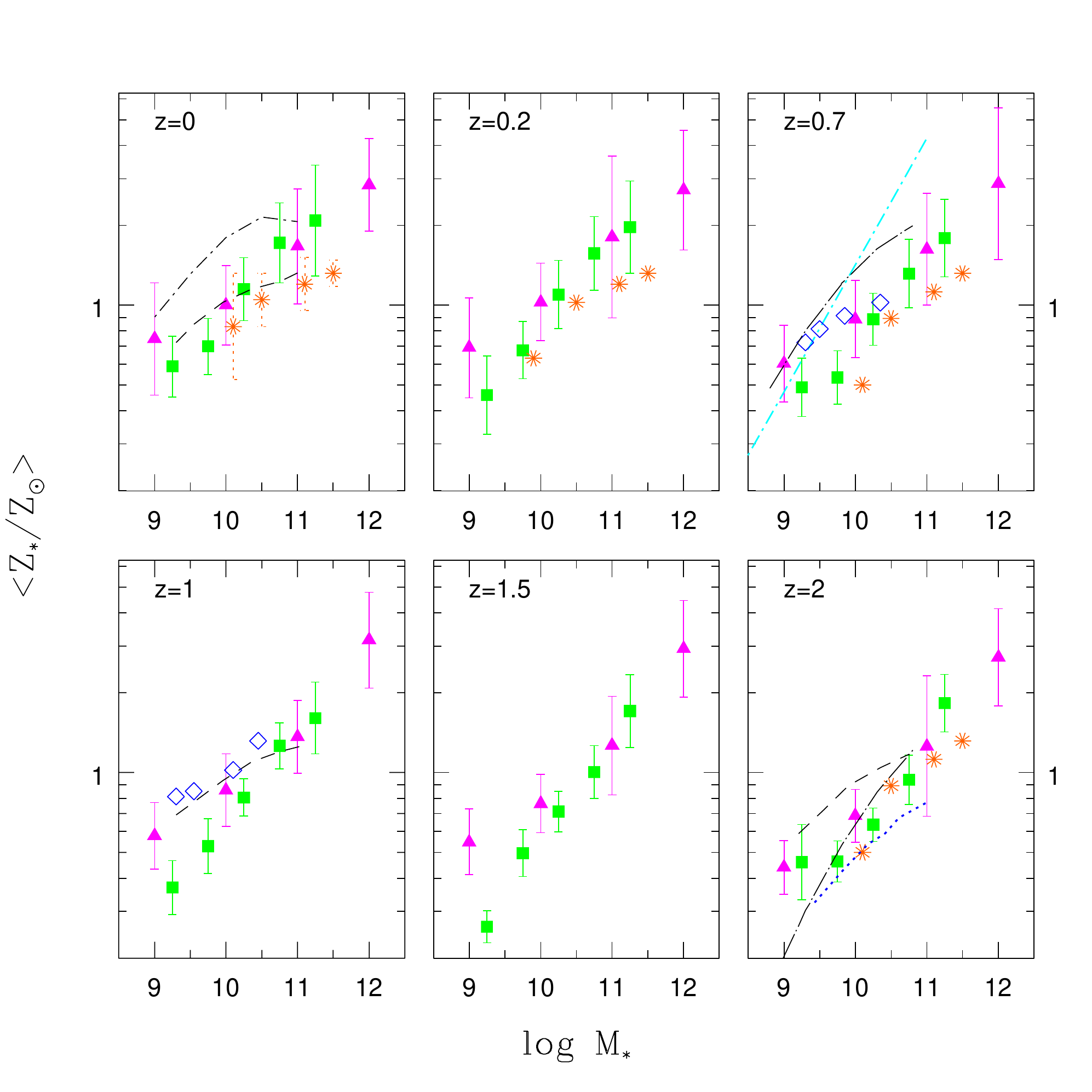}
\caption{Logarithmic relation between stellar (mass-weighted) global metallicity and stellar mass 
at different redshifts, 
for clusters (purple triangles) and groups (green squares). Here {\it all} galaxies are included. 
Errorbars give the scatter around the average metallicity within each mass bin. 
Observational points are: E09 from SDSS at $z\simeq$0.07 as dot-dashed curve (polynomial best-fit); 
S05 at $z\simeq$0.7 (linear fit over a redshift range 0.5-0.9) as cyan dot-dashed curve; PM09 at $z\simeq$0.7 and 1 as blue open diamonds;
Erb06 at $z$=2.2 as dotted curve; M08 at $z$=0.7, 2.2 as dot-long dashed curves (quadratic best-fit).
Also shown simulations from TDRS05 at $z$=0,1,2 (as dashed lines) 
and S.A.M. by VA09 (as orange asterisks; including all environments).
All data (except VA09) have been converted to mean metallicities by means of the
approximation $12+log(O/H)=log(Z/Z_\odot)+8.69$, with $Z_\odot$=0.0134 (Asplund et al. 2009).
Testing with luminosity-weighted metallicities has yielded very similar results.
}
\label{ZM}
\end{figure}

\section{Results}

\subsection{Evolution of the SF efficiency}

It is noteworthy to remind here that all our galaxies follow on average a SFR evolution such that the star forming activity is steeply declining since $z$=2
(see Romeo et al. 2005), consistently with Madau plots for both cluster and field galaxies.
To this respect, Figs. \ref{SFR} and \ref{sSFR} show the SFR and sSFR as a function of stellar mass, respectively, at different redshifts, where
the threshold introduced in Sec. 3 is evidenced (dotted lines).
First of all, the SFR-mass correlation and the sSFR-mass anticorrelation are both fairly reproduced. Our galaxies are compared with the well-established relations observed 
at $z$=0, 1 and 2, from SDSS and GOODS data for star-forming galaxies: good concordance can be evidenced, except that at $z$=2, where the simulated active galaxies lie below
the observed average. Apart of the normalization, the evolution of the two galaxy populations can be followed across time: a passive sequence of
low-metallicity galaxies starts to being built at least from $z$=2 and from the low-mass side; higher-metallicity and higher-mass objects begin to fall down 
onto this passive sequence just after $z$=2, filling the mass gaps until completing the horizontal sequence at around $z$=0.7.
At the same time the active galaxies that have not yet fallen onto the passive sequence, tend to form a tighter, although lesser populated, relation in the sSFR-mass plane;
in the SFR-mass plot instead their scatter increases steadily with time, until the complete distruction of the original tight sequence.
Next step to understanding this insight would be to extend such analysis to epochs beyond $z$=2, in order to better discriminate the birth epoch of the SFR-mass sequence. 

From the combined analysis of Figs. \ref{SFR} and \ref{sSFR} one can confirm the conclusions of R08, that is the RS follows the same consolidating evolution of the sSFR,
while parallelly the SFR--mass sequence breaks down with time.
In general our results are compatible with the S.A.M. described in Lilly et al.(2013), where the sSFR of the ``Main Sequence" of star-forming galaxies declines weakly with stellar mass as $sSFR\propto M_*^{-1}$, maintaining a tight dispersion at around 0.3 dex, and at the same time strongly evolving with redshift as $sSFR\propto (1+z)^3$ at least up to $z\simeq$2.

In Figs. \ref{mu_SFR} and \ref{mu_sSFR} the stellar metallicity is plotted against the new parameter $\mu_{min}$ proposed by Mannucci et al. (2010) as a
combination of both $M_*$ and $SFR$ that is alternative to sSFR and that minimizes the relation's dispersion; their own polinomyal fit (dashed line)
is drawn along for comparison to our points.
Contrarily to the universal (up to $z\simeq$2.5) validity of such a fit, we do find much more evolution in the star-forming galaxy population across time, especially in their scatter.
Our star-forming galaxies follow a defined trend increasing with $\mu_{min}$, with a scatter initially high at the low-mass end and eventually, after a turnpoint of maximum
scatter at $z\simeq$0.5, increased at high masses. 
Not shown in the figure are the completely passive galaxies, that occupy the bottom-right corner in each plot,
without any evolution besides of the growing number of members with time.
The metallicity generally increases with $\mu_{min}$, but interestingly the nature of such a trend varies whether the galaxies are furtherly classified according
to their SFR or sSFR: in the first case (Fig. \ref{mu_SFR}) the more active are also metal-richer, whereas in the second case (Fig. \ref{mu_sSFR}) the 
metal-richer are those with lower sSFR, that is lower efficiency in gas-to-stars conversion. 
Again this is in agreement with the simple analytical model developed by Lilly et al. (2013), where metallicity decreases as sSFR increases.

\subsection{The global ZM relation}

In Fig.\ref{ZM_SSFR} the ZM relation is plotted of both classes of galaxies is plotted, coloured according to their sSFR:
at any epoch the galaxies with higher sSFR are those at intermediate mass around $10^{10}M_\odot$, but a definite trend is
established at high redshift only above this mass, recovering that mass bimodality in the ZM slope already reported in
several works (see below). The SF galaxies (here: all but the red-coloured ones) smooth their slope with time, until merging
with the DS at low $z$. The latter is well established at $z\simeq$0.7 and presents a high scatter at low masses, probably
due to an excess of over-metallic dwarfs. At high masses instead the ZM relation is definitely shaped as a sequence in sSSFR,
namely in star formation efficiency.

In Fig.~\ref{ZM} we plot the stellar (mass-weighted) total metallicity as a function of mass, calculated as average
value in each mass bin, at different redshifts, for both cluster and group galaxies, including DS+SF.
The overall trend is approximately linear in the space ($log Z$, $log M_*$).
However we notice that the relation gets steeper with redshift at the bright end in clusters, whereas at the faint end in groups: 
the higher-mass galaxies in clusters and conversely the lower-mass ones in groups are mostly responsable for steepening the relation 
since redshift 1,
blending it towards either metallicity extreme, respectively.

VA09 have estimated the lookback evolution of local galaxies by applying a spectral synthesis 
code on SDSS data at $z$=0.1 (within the closed box approximation): a good overall agreement can be found with
their points, except at the faint end ($M\lsim 10^{10}M_{\odot}$) when they report a very steep relation that goes out of scale in both
our figures 1 and \ref{sloDS}$b$ (see below).
Simulations from TDRS05, also without outflows, are plotted too:
their mean metallicity decreases with $z$, but the trend with mass is shallower than ours and also than that
by VA09, especially at the faint end at high redshift, where the absence of galactic winds overestimates metallicity in low-mass galaxies.

As for observational data, our results fairly match those by PM09 at $z\simeq$0.7 and 1 over two orders of magnitude in mass,
while lie well below data from E09 at $z$=0.07 from SDSS.
The ZM relation from S05 at $z\simeq$0.7, calculated as a linear bestfit over a mean redshift, has a quite steep slope,
leaving lower all other data (including ours) at high-mass end; moreover, they do not find any significant evolution with respect to the two redshift halves
of the whole sample.
At $z\simeq$2 our ZM relation finds good accordance with VA09 at high masses and with Erb06 and M08 at the low-end. 
As to the slope at high redshift, VA09 and M08 present a much steeper relation with respect to Erb06 and TDRS05.
In particular, the curvature of our ZM relation would be flattening at high mass end if excluding the highest mass bin (corresponding to the CDs),
which is compatible with the quadratic best-fit proposed by M08 (see panel at $z$=0.7) and E08 (at $z$=0).

However, the comparison with observational data and even among them is not straightforward:
the slope evolution of the ZM relation is modelled in observational works as an extrapolation of redshift-limited samples,
under some strong assumptions such as the closed-box model (S05), or by assuming an unchanged shape along with a joint shift in mass and
metallicity (M08). In particular the latter authors adopt the same empirical function of the local ZM relation by simply changing a few parameters in
it, in order to provide the bestfit to the observed data at other redshifts.
In any case, it is worth noticing that there is no observational counterpart to our BCG mass bin, that stands as outlier with respect to the
extrapolated best-fitted quadratic curve at epochs $z\geq$1.

\begin{figure}
\includegraphics[height=8cm,width=8.5cm,clip]{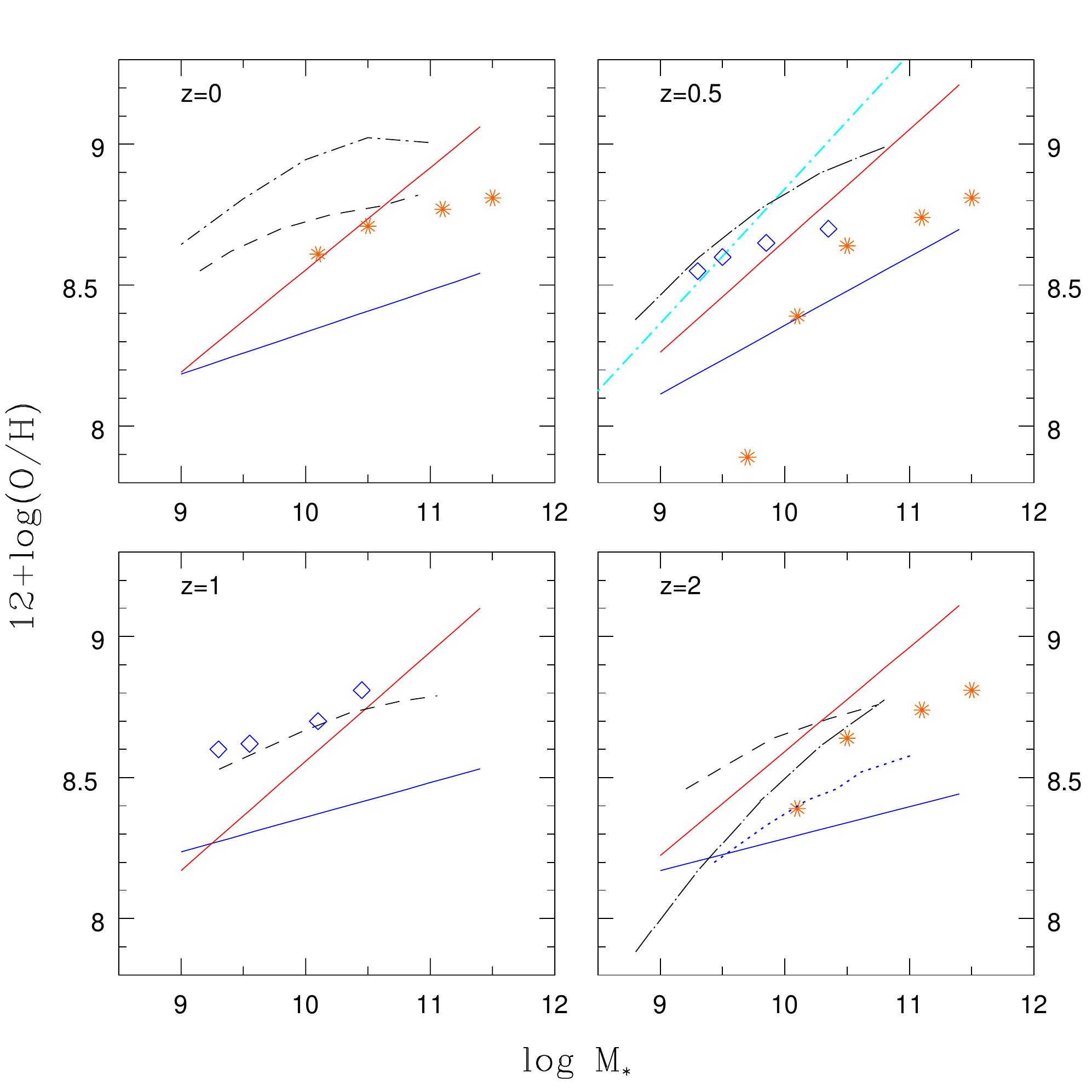}
\caption{Mass-weighted oxygen (blue) and total (red) gas metallicity--stellar mass relation, as linear best-fit for one cluster,
excluding the CD. A minimum threshold of gas particles has been applied. Data symbols as in previous figure.
\bigskip
}
\label{ZMgas}
\end{figure}

\begin{figure}
\includegraphics[width=8.5cm,clip]{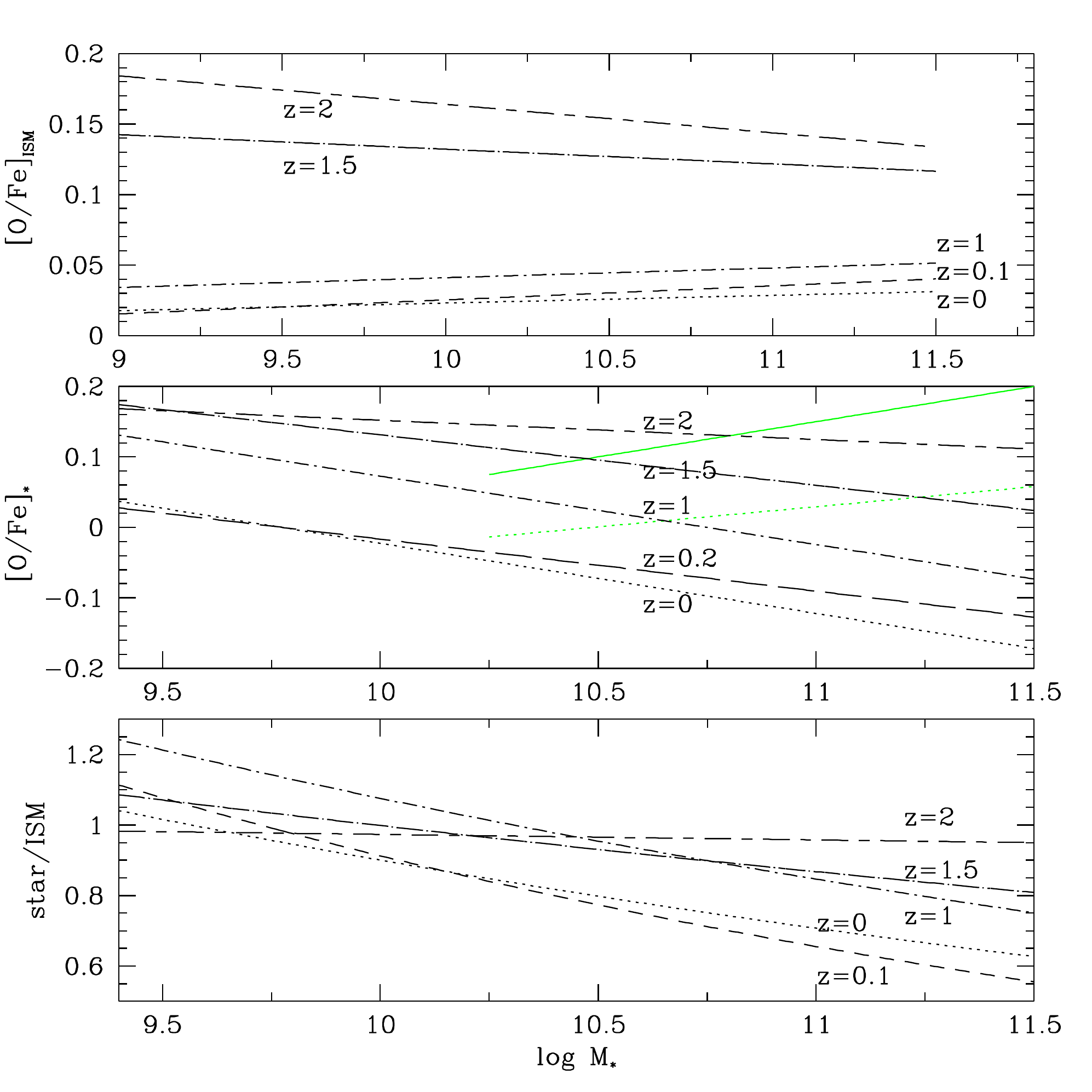}
\caption{Linear best-fits to the gaseous ({\it upper}) and stellar ({\it middle}) $\alpha$/Fe--stellar mass relation at different redshifts, 
for the same cluster as in previous figure, excluding the CD. 
The stellar $z$=0 (dotted) line is
split for the passive and massive galaxies only ($M_*>2x10^{10}M_{\odot}$, green) and compared with data from Thomas et al.(2007)
for a SDSS luminous-limited sample of nearby non star-forming ellipticals (green solid). The stellar fits are calculated over a
mass-limited sample at $M_*>$2.5$\times 10^{9}M_{\odot}$, below which the scatter is too high for a relation to exist.
{\it Bottom}: the ratio between O/Fe in stars and in the ISM is plotted. }
\label{OFe}
\end{figure}

\subsection{Gas abundances and $\alpha$/Fe}
 
In Fig.~\ref{ZMgas} we have plotted also the gas phase oxygen and total abundances, for the lesser cluster
as reference, excluding the BCG. Here our model attain to two main results:
first, the total gas abundance is at all epochs steeper than the oxygen one; and second, both show inappreciable evolution in
either their slope or normalization. 

In Fig. \ref{OFe} the $\alpha$/Fe relative abundance is plotted as a function of stellar mass for the same cluster at various epochs.
Here we get a flatter ratio than observed at low $z$.
The overall trend is slightly increasing, and with virtually no change in slope, up to $z$=1. 
On the other hand, the [O/Fe] inverts to decreasing at $z$=1.5 and 2, when 
it also gets on average higher by a factor two and three, respectively.
As a general feature, a decreasing of the relative abundance with time at given mass is expected as long as star formation keeps converting
iron enriched ISM into stars, what in turn makes the initially high abundance ratio to lower (see de la Rosa et al. 2011).

An increasing [$\alpha$/Fe] ratio with stellar mass in ET galaxies generally implies that more massive ellipticals formed earlier and 
faster than less massive ones and can be derived within a model with either star formation
efficiency as increasing function of galaxy mass, or with a mass-dependent IMF, top-heavier in more massive galaxies (Matteucci 1994).
Calura \& Menci (2011) ascertain that it is AGN late quenching of the starburst activity that can effectively account for the observed 
$\alpha$/Fe ratio in ellipticals.

Yet, it is a meaningful indicator on itself about the contribution of SN feedback to the total
$\alpha$-enhancement of the ISM: under this respect the increasing trend with galaxy mass at low redshift means that simple stellar ageing 
prevails in more massive galaxies upon the SN channel in releasing $\alpha$-elements, whilst in less massive ones the latter is still the
predominant mechanism in enriching the surrounding gas: from this comes the higher iron abundance relative to oxygen in smaller local
galaxies. This picture is consistent with less massive galaxies being site of enduring star forming activity, hence of SN-II episodes
bringing forth starbursts, than more massive ones -provided that the SN-Ia rate be of the same order across the two regimes.
Arrigoni et al. (2010) support that a correct $\alpha$/Fe--mass relation can only be reproduced if a top-heavy initial mass function and a low fraction 
of binaries that explode as SN-Ia are assumed, which are both prescriptions adopted in our model.

At high redshift ($z>$1) the trend overturns, consistently with a significant population of massive galaxies still in course
of active star formation at that epoch.

In the second panel the stellar [O/Fe] is plotted too: here a tight relation can be established only from $M_*>$2.5$\times 10^{9}M_{\odot}$, 
below which the scatter is too high to be compatible with a mass functionality. Under these limitations, the reported trend is here decreasing,
with a smooth flattening with redshift. 
The stellar [O/Fe] is lower than the ISM one at almost any mass for
$z\leq$0.2, that is the distribution of $\alpha$-elements with respect to iron is skewed towards the gaseous phase. 
At $z$=2 the ratio between the two is basically independent of the galaxy mass and is also fairly equiparted; since then,
progressively lower mass galaxies present higher metal content in stars than in the ISM, with the opposite for high mass ones --being $logM_*\simeq$10.3 the
turnover mass at $z$=1-1.5 (bottom panel).

The $z$=0 best-fit is also presented considering the more massive galaxies ($M>3 \cdot 10^{10}M_{\odot}$) only, to be compared
with data synthesized from a SDSS luminous sample of passive ellipticals (Thomas et al. 2007): both converge in giving a much steeper relation, result that is
also concord with a semi-analytical model by Pipino et al. (2009). In particular our fit reproduces the same slope as observed in that mass regime, 
although at considerably lower values of [O/Fe]: the reason for this may root into an excess of $\alpha$-enhanced satellites, as previously pointed
out by, again, Pipino et al. (2009). 


\begin{figure}
\includegraphics[width=8cm,clip]{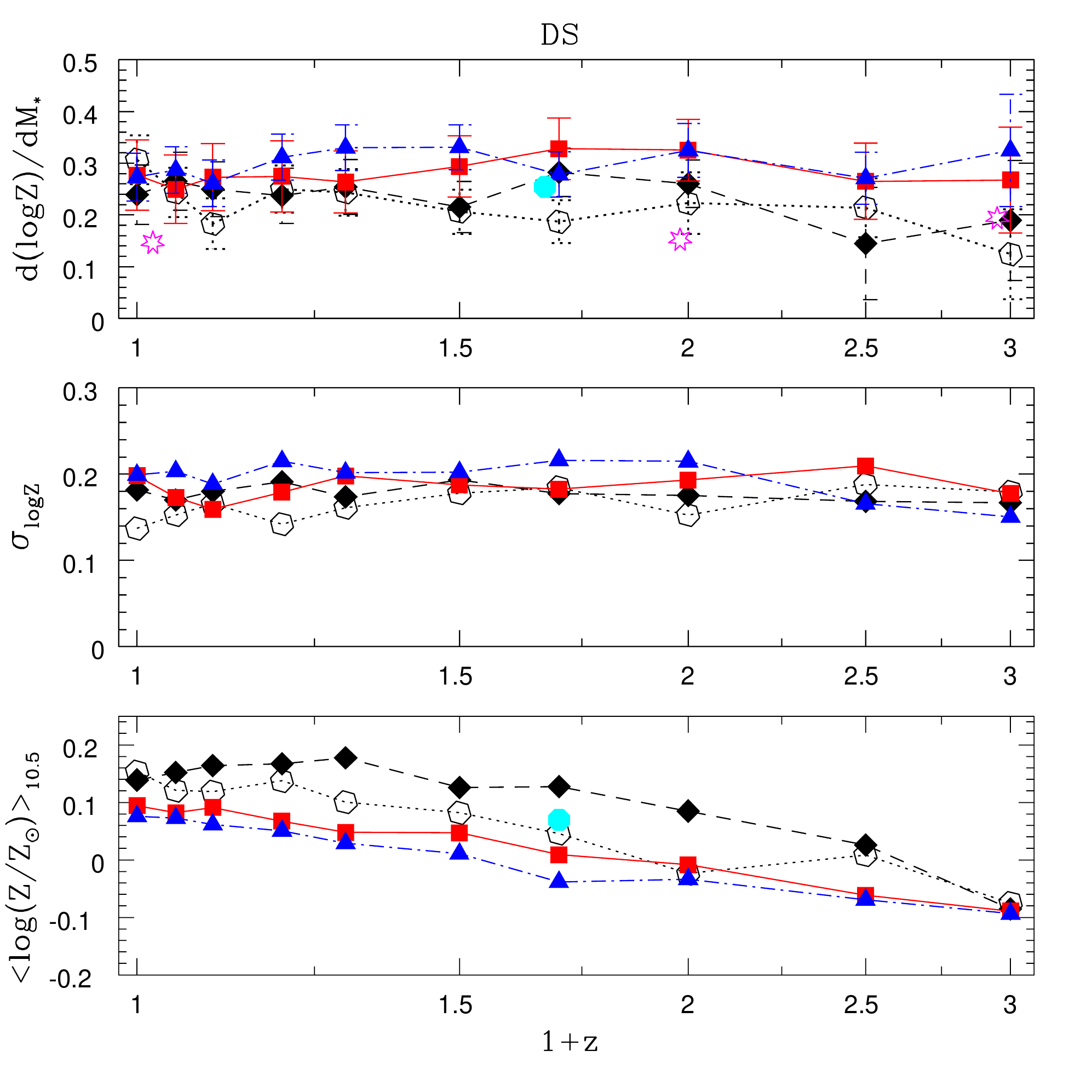}
\includegraphics[width=8cm,clip]{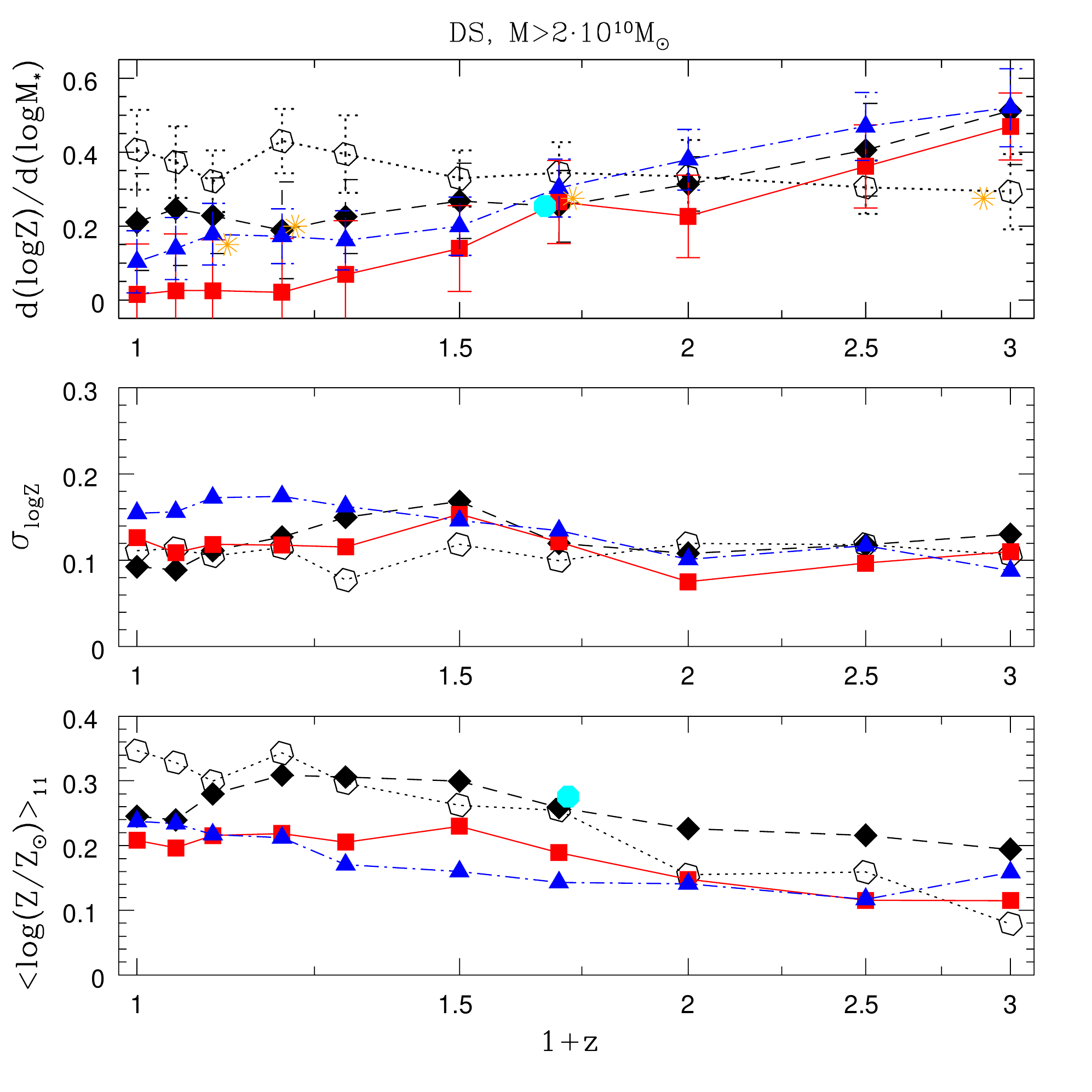}
\caption{{\it a)} Slope, scatter and characteristic mass (calculated at $logM_*$=10.5) of the ZM relation
for DS galaxies: filled diamonds for cluster cores, 
open circles for cluster outskirts, blue triangles for normal groups, red squares for fossil groups.
In the upper and bottom panel we compare with simulations from TDRS05 (magenta open stars)
and data from S05 (cyan solid circles).
Error-bars indicate the Poissonian error at each redshift and are only drawn for the slope for simplicity.
Pairwise, overlapping data points are slightly offset in redshift for clarity's sake.
Here, a completeness limit of $M_*> 10^{9}M_\odot$ has been adopted (see R08 for details).
{\it b)} Same as above for galaxies more massive than $2\times 10^{10}M_\odot$ (same symbols).
Data points for the slope are from VA09 (orange asterisks, averaged over all environments)
and S05 (cyan solid circles).}
\label{sloDS}
\end{figure}

\begin{figure}
\includegraphics[width=8cm,clip]{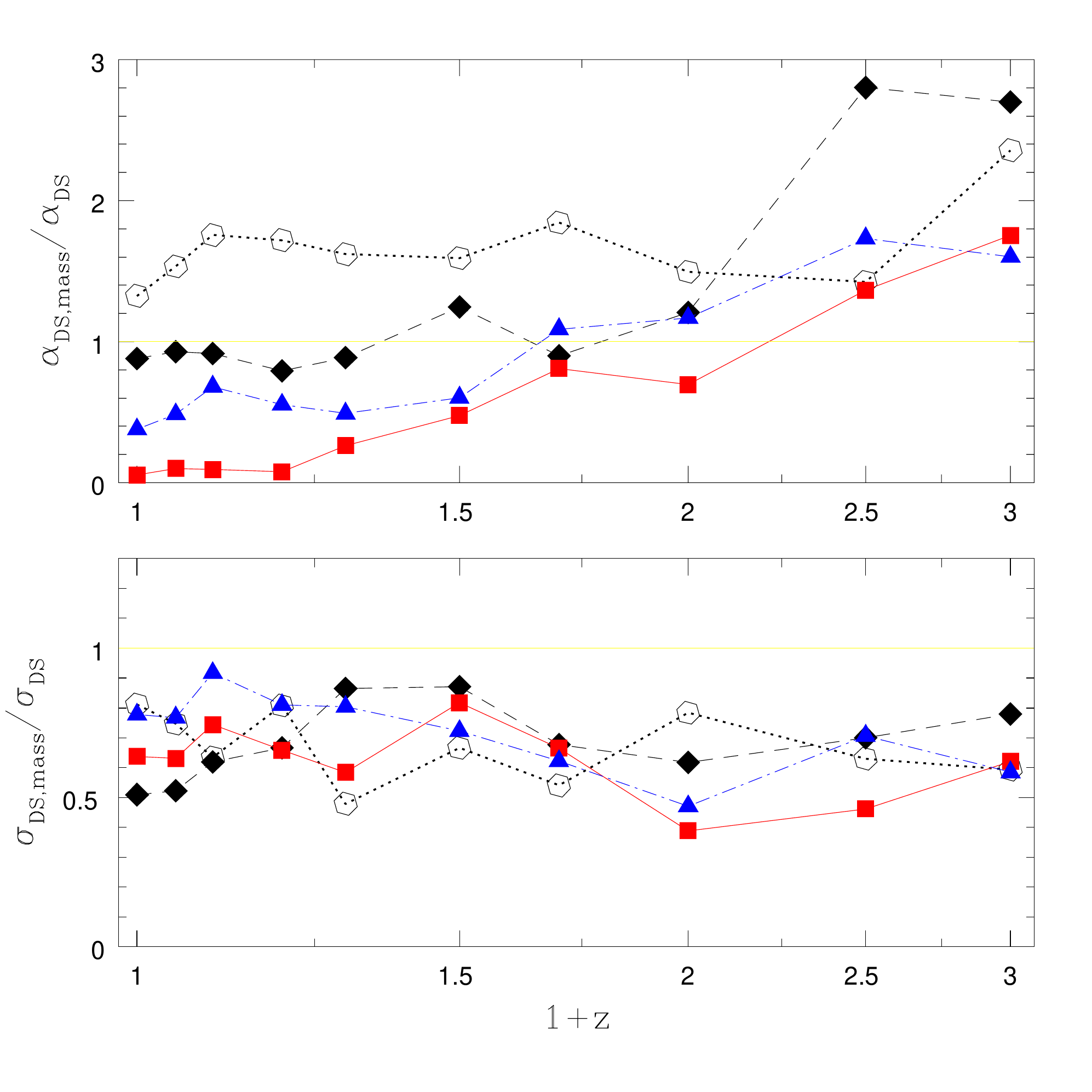}
\caption{Ratios of the slope and scatter of the ZM relation for the massive DS to overall DS sample.
}
\label{ratDS}
\end{figure}

\begin{figure}
\includegraphics[width=8cm,clip]{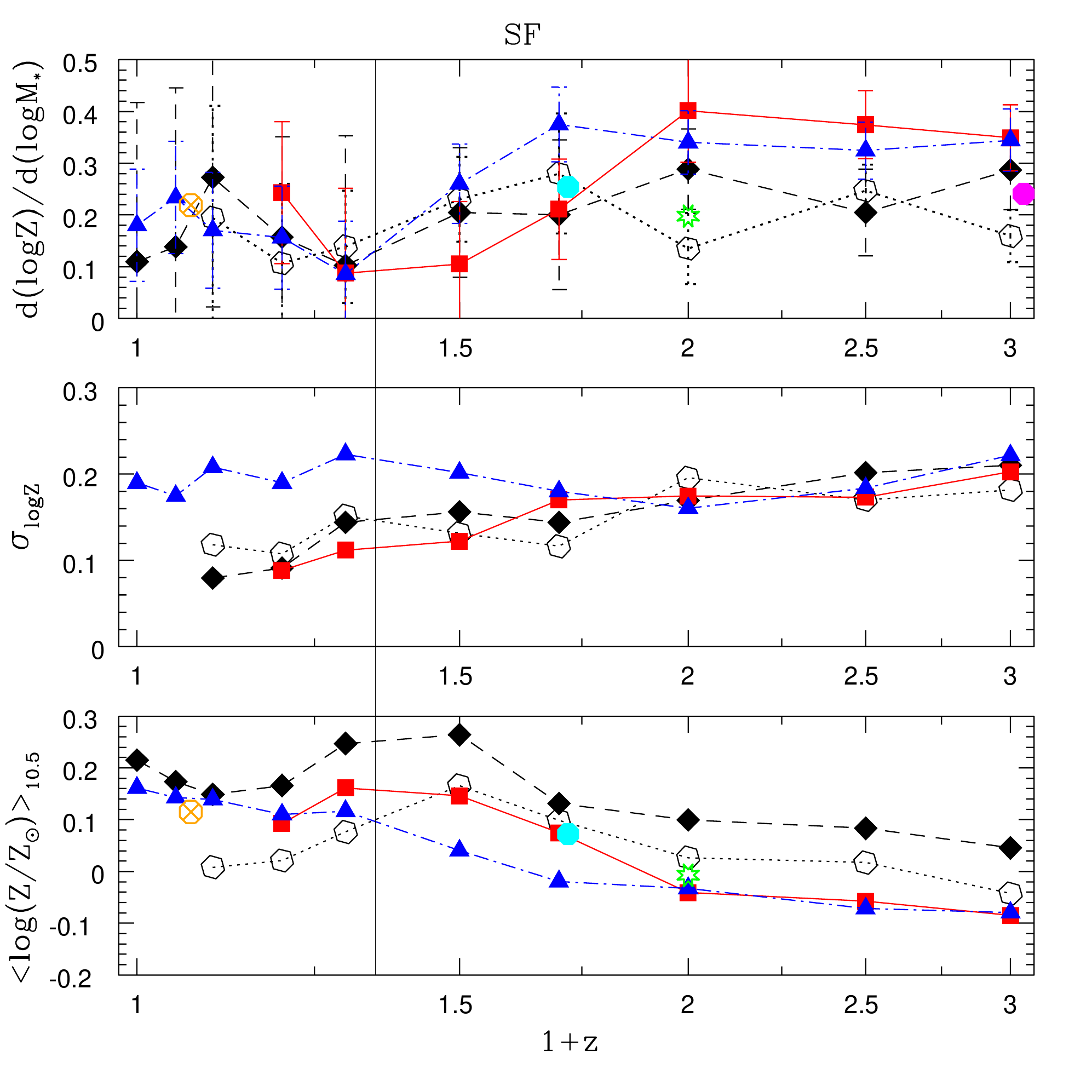}
\includegraphics[width=8cm,clip]{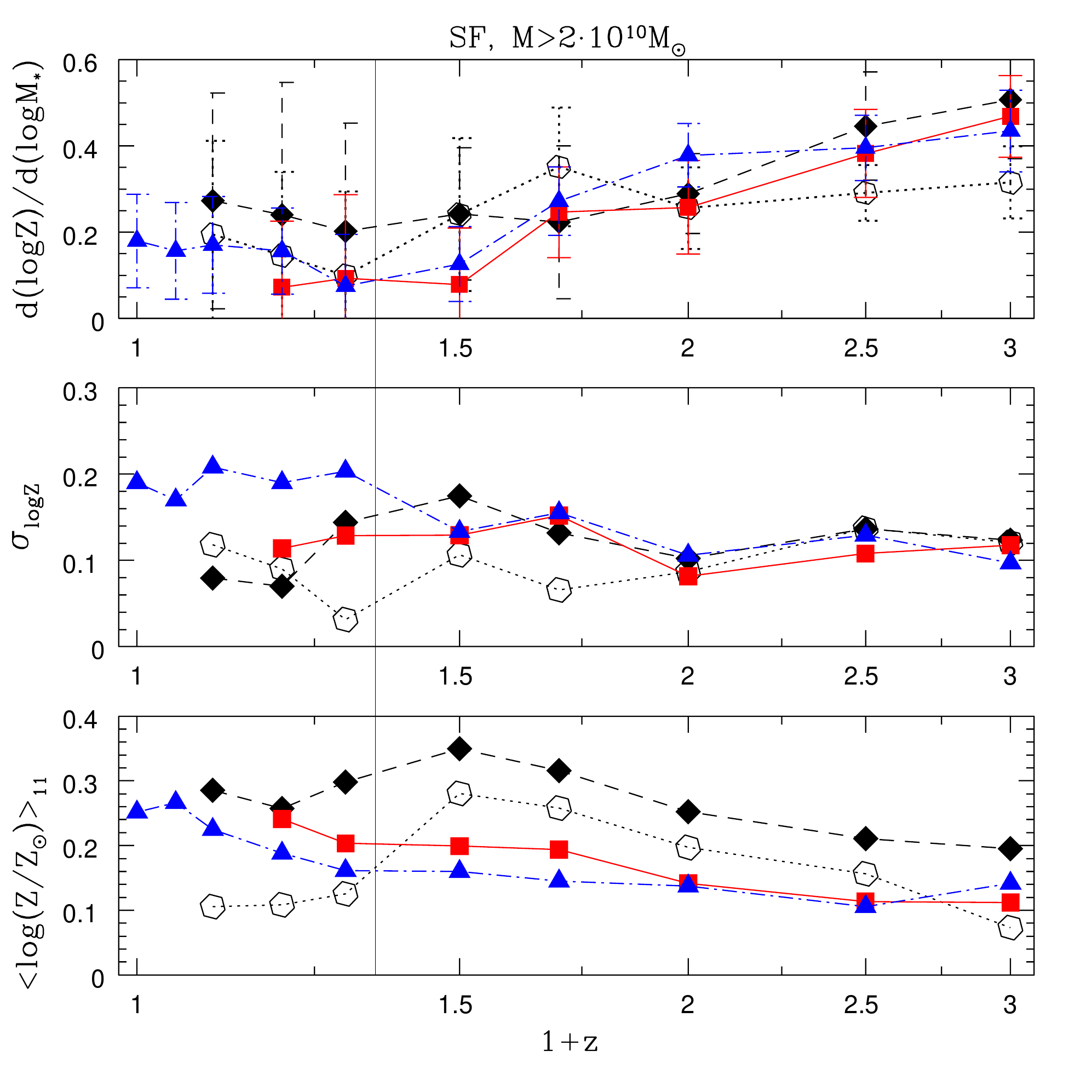}
\caption{Same as Fig. \ref{sloDS} for star-forming galaxies. The vertical line indicates the confidence limit in redshift, below which the number of fittable 
objects is too low as resulting in the Poissonian error. Data points: PM09 (green stars), Erb06 (magenta solid circles),
Kewley \& Ellison (2008: orange crossed circles), S05 (cyan solid circles): all except the latter are star-forming samples. 
}
\label{sloSF}
\end{figure}

\begin{figure}
\includegraphics[width=8cm,clip]{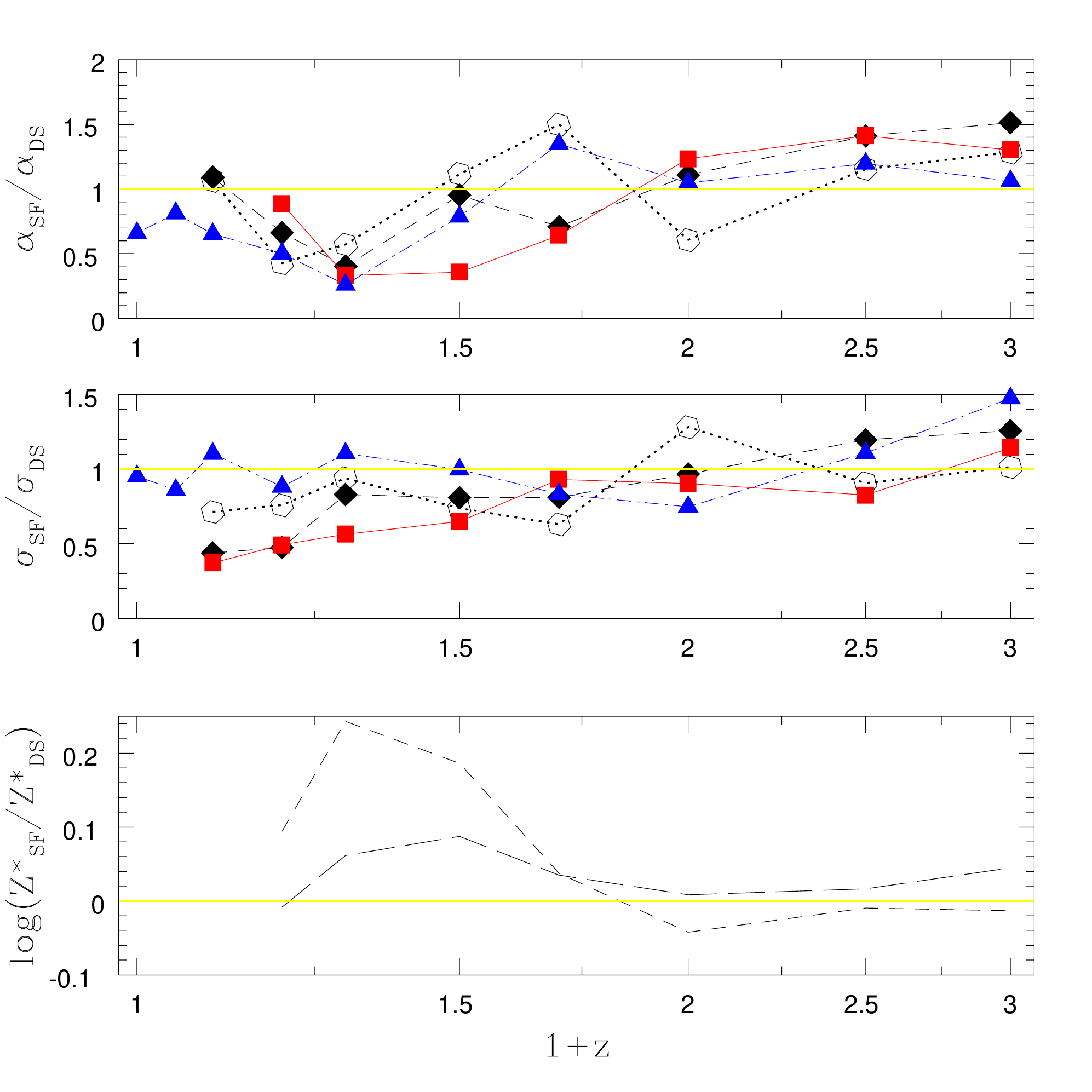}
\caption{Ratios of the slope (upper panel) and scatter (middle) of the ZM relation for the SF to DS sample.
Bottom panel: ratio of the characteristic metallicity at $log M=9.5$ (short-dash) and $log M=10.5$ (long-dash) 
for the SF to DS sample, averaged over all environments.
}
\label{ratSF}
\end{figure}

\subsection{Slope evolution: The passive sample}

We derive the slope of the ZM relation from the logarithmic relation plotted in Fig.~\ref{ZM},
as $\alpha_{ZM}= \delta(logZ)/\delta(log M_*)$,
which is equivalent to a power law $Z \propto M_*^{\alpha_{ZM}}$.
Its evolution with redshift is plotted in Figs.~\ref{sloDS} and \ref{sloSF}, along with
the scatter and zero-point (or intercept at $logM_*=10.5$) of the ZM relations calculated in the different samples. 
These plots further confirm the previous results outlined above for the overall galaxy population in Fig.\ref{ZM} .

As a general feature, we first note that in all the cases studied the amplitude of the ZM relation (bottom panels) smoothly increases with time, 
confirming that earlier (less evolved) galaxies, presumably with higher gas fractions, have also lower gas-phase metallicities (cfr. S05, Erb06), 
which translates into lower stellar ones as well.
Moreover, cluster core galaxies have a higher metallicity than in groups and than cluster outer galaxies, irrespectively to their star
forming activity, and at all redshifts: this is in accordance with E09, 
who find an offset of $\sim$0.03 dex between inner and outer galaxies in clusters. 

For what concerns the DS, results are displayed in Fig.~\ref{sloDS}: 
we find that the slope of the ZM relation remains roughly constant with redshift, 
at average values between $\sim$0.2 for clusters and 0.3 for groups, with fluctuations more evident at higher redshift compatible
with the Poissonian errors (which for passive galaxies increase with redshift).
Its scatter maintains around 0.2 almost constantly with time, and the mean metallicity decreases with redshift for all classes.

In particular, cluster cores stay at super-solar level throughout up to $z\sim$1.5, with an offset respect to cluster outskirts
comparable to that by E09. 
Although not explicitely separated in the plot, higher mean metallicity is presented by the larger, less dynamically relaxed of the two clusters, 
both in the core and in the outskirts. 
Moreover, a sequence can be drawn in the characteristic metallicity (third panel), going from cluster IN through OUT, FG and NG, that is the same environmental
sequence found in R08 for galaxies to complete their approach to the RS.

When looking at DS massive galaxies only ($M_*>2\times 10^{10}M_\odot$, Fig.~\ref{sloDS}$b$), the main result is that
the ZM slope increases steeply with redshift (with the exception of cluster outskirts), indicating that 
the bulk of its evolution is mostly driven by more massive galaxies. 

As by comparing Figs.\ref{sloDS}$a$ and \ref{sloDS}$b$ (with the aid of Fig.\ref{ratDS}, showing the slope and scatter ratios between the two samples), 
also the contribution to the slope is mainly
given by the massive systems at $z\gsim$1, when it starts to deviate from that of the overall DS sample.
Only in the outer cluster galaxies, the massive subsample contribution to the total slope  is overwhelming at all epochs,
following a quasi-constant pattern: that is, massive outer cluster galaxies have a steep distribution in the ZM plane
with little evolution in time.
Opposite behaviour is observed in massive group galaxies, which have a flat low-$z$ ZM relation (especially FGs), but with a 
strong redshift evolution of the slope thenceforth.

In Fig.\ref{sloDS}$b$ we also compare ours with the slope data points from VA09, limited to the massive end ($M>1.5\times 10^{10}$) and averaged 
over all environments: they closely follow those of our cluster cores up to $z\sim 1$.
These authors indeed found a slight increase of the ZM slope up to $z$=1.9, when only considering the massive subsample; 
on the less massive side instead, the reported slope was much
steeper, reaching extreme values of 1.2 at $z$=0.7, up to 1.5 at $z$=1.9 -what leads to a very strong mass bimodality
quite at odds with any other data.

In both Figs.\ref{sloDS}$a$ and \ref{sloDS}$b$ no significative evolution of scatter is reported;
for massive galaxies, it remains almost constant again, yet at a value between 0.1 and 0.15, which makes more than 0.05 dex
lower with respect to the whole DS sample -namely, the scatter of the ZM relation for passive galaxies is dominated by low-mass
systems. 

\subsection{The star forming sample}

In Fig.~\ref{sloSF}a we analyse the SF sample, with the caveat that only results from $z\sim0.3$
onwards can be considered significant, given the low number of SF galaxies found at lower redshift.
Here we compare with some of the many observational data available for star-forming samples at different epochs:
their slopes at $z$=0.1, 0.7, 1 and 2 are consistent with our points for clusters, although draw a quasi-constant
evolution altogether. Indeed, our slope here evolves more rapidly than in DS galaxies up to $z\sim1$, after which it gets almost constant too,
and again higher in groups (first panel).

As for the scatter, it is increasing with redshift up to beyond a value of 0.2 (second panel).
This is expected, since this is the same galaxy population of the {\it blue cloud}, reducing its scatter throughout their approach towards the RS (see R08).

As for the characteristic metallicity at given mass (third panel), in the galaxy cores it reaches a maximum 
at $z$=0.5, thence decreasing along a hump shaped curve: this reproduces to a very good extent the trend found by PM13,
who pointed out a strong acceleration at increasing the oxygen abundance before $z\sim$0.5, somewhat braked thereafter.
Moreover, the metal abundance was around half the present value at $z\sim$1 (average over all classes), that is consistent again 
with findings by these same authors.

In Fig.~\ref{sloSF}b the contribution to the slope, scatter and normalization by SF $M>2\times 10^{10}M_\odot$ galaxies only is analyzed.
As in the DS, the slope evolution is dominated by more massive systems for $z\gsim$0.5-0.7, except in outer cluster regions where the ZM relation
steepens very slightly with redshift.
Also the scatter keeps basically constant and at lower level than in whole sample, indicating that it is mostly driven by smaller galaxies 
of the {\it blue cloud} at earlier epochs.

Finally, we compare in Fig.\ref{ratSF} the respective ZM properties of both the DS and SF samples. 
For what concerns the slope (first panel), the ZM relation keeps slightly steeper in SF than in DS
until $z$=1 (IN), 0.8 (FG), 0.6 (NG), that is consistent with the epoch when the SF galaxies complete their migration towards the RS:
i.e. z$\sim$1, 0.7 and 0.5 for IN, FG and NG respectively, as from R08.
Similar trend is followed by the scatter, that is mostly proceding by the SF population at higher redshift than about 1, 
whilst in cluster cores and FG at low $z$ that of the DS sample is prevailing (second panel); SF galaxies in NG present higher scatter
at almost all epochs.
Such trend simply follows that of $N_{pas}/N_{act}$ as outlined in R08.

Regarding the ratio of mean metallicities between the SF and DS samples (third panel), it maintains about unity until $z\sim$0.7.
Afterwards, it undergoes an uprise, especially marked when calculated at $logM=9.5$, bringing it
at values higher than unity within a redshift range from 0.7 to 0.2.
Such trend is explained by the slope's behaviour in the first panel:
at nearer epochs than $\sim$0.7, DS galaxies have a higher slope than SF ones, yet lower absolute metallicity; 
this results into a reduced difference between the mean abundance at logM=10.5 with respect to that at logM=9.5. 
Reversely, for $z\gsim$0.7 the SF galaxies acquire a steeper ZM relation, while DS ones keep their almost constant slope:
this gives rise to a positive $Z_{DS}-Z_{SF}$ at the lower mass bin, while negative at higher mass.

These results are in partial accordance with Yates, Kauffmann \& Guo (2012), who observed and modelled at $z$=0 a low-mass galaxy population
supporting an anti-correlation between mean (oxygen) abundance and SFR, whereas in the high-mass regime less active galaxies are metal poorer; 
though their threshold in mass is logM=10.5, which gives little room for our high-mass end systems in the ZM plane. 
They explain this trend by massive local metal-poorer galaxies having undergone a gas-rich merger
in the past, triggering a starburst that exhausted most of the cold gas, leaving little to further star formation: in this scenario the 
turnover in the ZM relation would then be due to the dilution of the gas phase metallicities in these systems. 
The inversion in the ratio we find at $z\sim$0.7 can be due to similar mechanisms
acting in our cluster galaxies as a delayed response to the major merger they experienced at $z\sim$0.8-1.5, when bearing in mind that the SFR parameter as 
here defined traces the past star forming activity over approximately the last Gyr.
Pairwise, Lara-L\'opez et al. (2010) reported a star forming population in SDSS galaxies, mostly composed by late-type, {\it blue cloud} galaxies, whose abundance
is positively correlated with SFR, corroborating our findings at $z\lsim$0.7.

In any case it is worth stressing that we are here comparing two classes of galaxies whose chemical properties are observable by means of 
different diagnostic tools. In particular in this subsection we are not focussing on the dependence of metallicity on SFR or SSFR, that has been previously
shown in Figs. 3 and 4 for the star-forming only sample; rather we are here looking at the average metallicity within each sample and at its
variation with stellar mass. 

\begin{figure}
\includegraphics[width=8cm,clip]{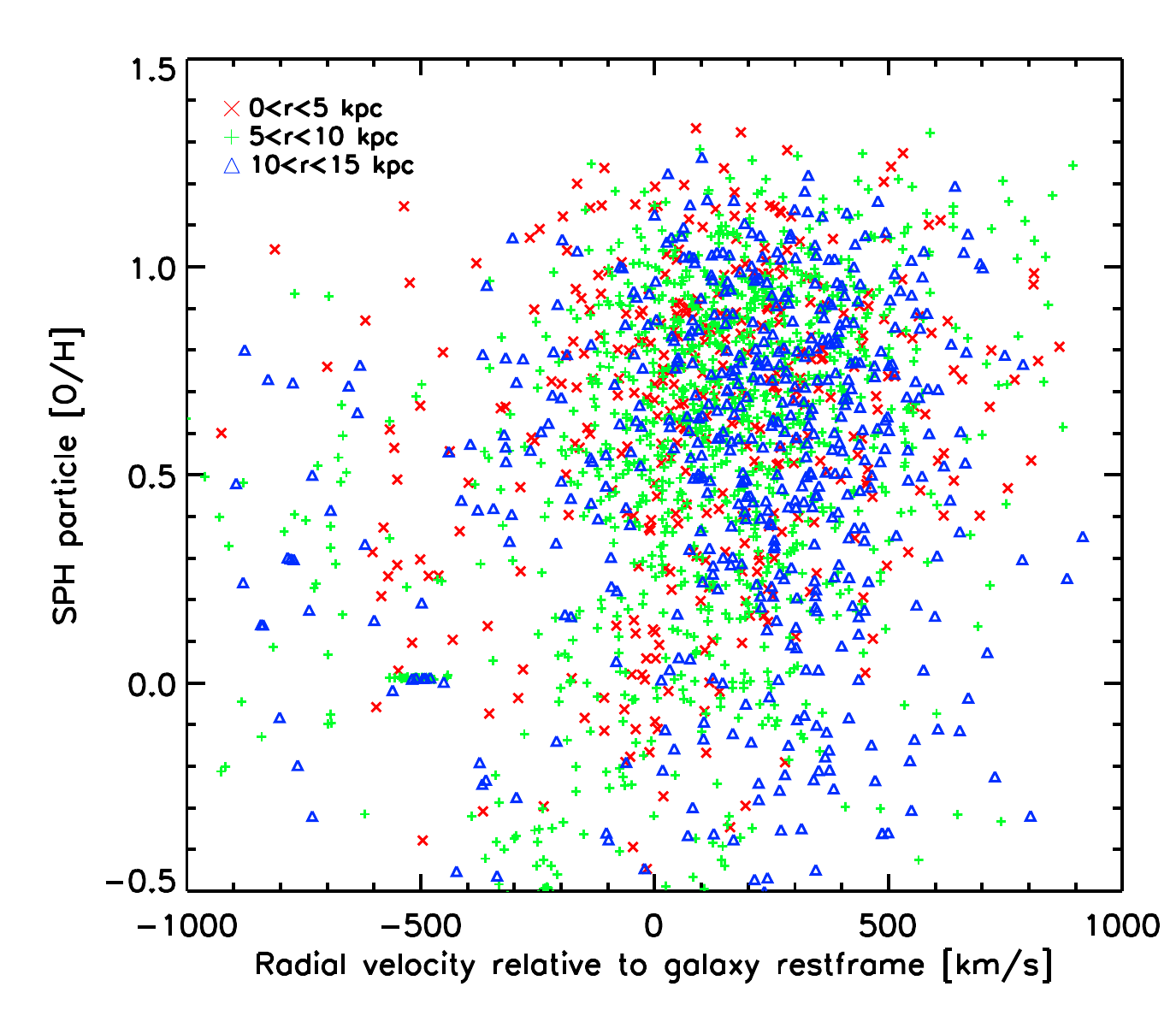}
\caption{Relation between [O/H] and radial velocity relative to the galaxy, for the SPH particles
within 15 kpc galactocentric distance. SPH particles within 0-5 kpc are shown by red crosses, 5-10 kpc by green crosses
and 10-15 kpc by blue triangles.  
\bigskip}
\label{outflow}
\end{figure}

\subsection{The role of galactic winds}

In order to assess the effect of our super-wind scheme on the outflows contribution to the variation of galaxy metallicities, we 
selected the largest ($M_*\simeq10^{11}M_\odot$) galaxy in the {\it Virgo} simulation at $z$=3.1, that displays a clear asymmetric star-burst driven outflow.
Fig. \ref{outflow} plots [O/H] for the (mostly hot) gas particles
within 15 kpc galactocentric distance versus their radial velocity relative to the galaxy's centre. From this one can infer:

{\it a}) Most SPH particles are outflowing in this frame, \\
\indent {\it b}) The wind is highly metal-enriched with median [O/H]$\sim$0.7-0.8 and mean [O/H] even higher.\\

Identifying the gas particles belonging to the winds is not straightforward, since 
some of the particles with positive radial velocities may not be part of the outflow itself, but
just have dynamical motions. We tried with different thresholds that could yield reasonable definitions
of {\it outflowing particle}:
in particular, if one defines the superwind as particles of $logT>$4.5 and $v_{rad}>$+500
km/s, then for these particles we found that
the median [O/H]=0.65 and the mean oxygen abundance is 6.76 times solar.
Changing these values to looser ones of $logT>$4.5 and $v_{rad}>$+300
km/s, gives for their median [O/H]=0.53 and a mean oxygen abundance of 5.25 times solar.

Besides to this, we verified that in the same galaxy the cold gas ($logT<4.5$) with $v_{rad}<$0, that is potentially
star-forming gas not affected by the asymmetric outflow, has a median [O/H]=0.15 (mean [O/H]=0.45).
The young stars (age$<$34 Myr) responsible for producing the outflow have a
median [O/H]=0.12 (mean [O/H]=0.48), whereas the entire galaxy (all stellar ages) has an
average [O/H] about 50\% lower than this.

We can conclude that the outflow is clearly significantly metal-enriched compared to the
star-forming cold gas left behind in the galaxy, and also compared to the stellar component. 
We also extended such analysis to other galaxies at lower redshifts, confirming these main findings.

\section{Discussion and conclusions}

Our cosmological--SPH simulations  
aimed at modelling in a self-consistent way the chemical enrichment of the inter-galactic medium (IGM) in reply 
to the stellar feedback (SN and galactic winds).
The physical processes at work in the codes include:
metal dependent radiative cooling, star formation, SN feedback generating galactic super-winds, chemical evolution with not instantaneous recycling into the IGM. The combination of these modules allows to reproduce most of the phenomena involved in the galaxy formation -namely,
galaxy mergers, infalling of extra-galactic gas, outflows of enriched ISM, along with internal physics driven by stellar feedback.

We extracted galaxies belonging to different environments from simulated clusters and groups and analyzed their
ZM relation and $\alpha$/Fe dependence on $z$ and mass. 
These were further classified as star-forming (SF) or passive (DS) according to their sSFR, in order to assess 
the contribution of star formation efficiency in shaping the ZM relations and their variation. 
We also studied how the SFR ans sSFR vary with stellar mass, and besides to this the dependence of metallicity on
each of these parameters, following the threefold plane proposed by Mannucci et al. (2010).
In the following are listed the main results about the amplitude and evolution of the slope and scatter of the ZM relations,
and about the combination of the three parameters $Z$, mass and SFR (or sSFR);
results that are however dependent on setup choices such as the stellar (rather than gaseous) metallicity, the strength of galactic winds,
or the IMF (top-heavy rather than Salpeter) in our models:

\bigskip
1) The star-forming galaxies make up a tight sequence in the SFR-$M_*$ plane at high redshift, whose scatter increases with time alongside with the consolidation of the passive sequence (Fig.\ref{SFR}). 
At the same time an anti-correlation between sSFR and stellar mass can be deduced (Fig.\ref{sSFR}), according to
the paradigm of galaxy downsizing. 
Likewise, and relatively to the SF sample only, an anti-correlation between sSFR and metallicity can be established (Fig.\ref{mu_sSFR}) , 
while on the contrary more active galaxies in terms of simple SFR are also metal-richer (Fig.\ref{mu_SFR}).
These findings confirm the role of the sSFR as the most reliable tracer of the star formation efficiency.
Moreover, the ZM relation itself is shown to being built as a sequence in sSFR, at least at high masses, while
at low mass a high scatter prevails (Fig.\ref{ZM_SSFR}).

\bigskip
2) The overall stellar ZM relation gets steeper at $z\gsim$1, due to low-mass galaxies in groups and high-mass ones (BCGs) in clusters.
Yet when excluding the highest mass bin corresponding to the BCGs, its shape is compatible with quadratic functional forms estimated in
particular by M08 at different redshifts (Fig.\ref{ZM}).

\bigskip
3) The gaseous oxygen abundance presents very scarce evolution with redshift (Fig.\ref{ZMgas}); its dependence on mass emerges however as shallower
than observed at $z$=0, perhaps due to an insufficient non-stellar feedback in our modelling of late massive galaxies (see Pipino,
Silk \& Matteucci, 2009).

\bigskip
4) The (both gaseous and stellar) [O/Fe] abundance ratio supports a clear trend to growing with redshift at given mass, following the natural
iron enrichment of the ISM that occurs on longer timescales with respect to the SN-II ejected $\alpha$-elements associated with
early starbursts. Its expected increasing slope with mass is recovered provided that only more massive (than $3\cdot10^9M_\odot$)
galaxies are considered, though at lower level of stellar [O/Fe] than observed (Fig.\ref{OFe}).

\bigskip
5) In the DS sample, groups present on average $\sim0.1$ dex higher slope than clusters.
The summed up slope in groups is the result of a combined effect by high-mass galaxies at $z\gsim$0.7, and by low-mass at nearer epochs.
For cluster cores, the global slope is dominated by more massive galaxies from the same epoch $z\sim$0.7 onwards.
In cluster outskirts instead the whole contribution to the slope is given by high-mass galaxies at all redshifts, meaning that low-mass 
outer galaxies keep a quasi-flat ZM relation (Figs. \ref{sloDS} and \ref{ratDS}).

\bigskip
6) In the whole DS sample, little redshift evolution is observed for the slope in all environments.
On the contrary, the slope of the ZM relation is increasing with $z$ for high-mass only galaxies.
This does not hold for cluster outskirts, where no slope evolution is reported even in the high-mass subsample (Figs. \ref{sloDS} and \ref{ratDS}).

\bigskip
7) The scatter of the ZM relation in the DS sample keeps fairly constant and tight over time.
It is mainly dominated by low-mass galaxies, which lie on a $\sim$0.05 dex higher value with respect to higher mass ones (Figs. \ref{sloDS} and \ref{ratDS}).
The scatter in the SF sample instead is increasing with redshift and is higher than that of the DS at $z\gsim$1, indicating that
the relation is still under construction at earlier epochs (Figs. \ref{sloSF} and \ref{ratSF}).

\bigskip
8) The slope of the ZM relation presents stronger overall evolution in the SF sample than in the DS, being higher at $z\gsim$1 
and lower at closer epochs (Fig.\ref{ratSF}).

\bigskip
9) The characteristic metallicity (at log M*=10.5) decreases steadily with increasing redshift, in both DS and SF,
clearly following the accumulation of metals with the cosmic time.
This is confirmed as a universal behaviour, as much well established as the corresponding decline in SFR in all galaxies, since a peak at some
epoch slightly earlier than $z$=2 that is our farther redshift considered in the present analysis (see Romeo et al 2005).

At all redshifts, it is highest in cluster cores (concording with results by E09)
and lowest in NGs: this may be correlated with the later lasting star formation activity in the latter systems, as found in R08.
Moreover, SF galaxies in clusters follow an accelerating pace up to $z\sim$0.5, slowed down at nearer epochs.
The difference in metallicity between massive and faint galaxies is more marked in the DS sample, where it amounts to twice
at almost all epochs (Fig.\ref{sloDS}).

From the comparison between our two classes (Fig. \ref{ratSF}), we can evince that
the characteristic metallicity in SF galaxies taken at $log M^*$=10.5 is slightly higher than that of DS at all epochs;
when estimated at $log M^*$=9.5 instead, it is slightly lower for $z\gsim$0.7 for SF galaxies than in passive ones,
but with an inversion at around that epoch, after which it gets considerably higher until recent times.
Such trend for $z\lsim$0.7 is imputable to low-mass metal-rich galaxies contributing to flatten the slope of the ZM relation in the SF sample.

\bigskip
10) The effect of galactic winds triggered by SN events acts as to ejecting from the galaxy gas particles having
higher metallicity than those of the star forming gas and than the star particles as well (Fig.\ref{outflow}).
This translates into a lowered galaxy abundance and is presumably enhanced in lower-mass systems, where the outflows attain
higher radial velocities -ultimately contributing to differentially change the ZM slope.
\\

To summarize, we find a ZM relation that is basically determined in its shape and evolution by its parametrization in terms of sSFR,
that is ultimately function of star forming efficiency (SFE). Moreover, a stronger evolution is reported for the metallicity of
more massive and passive galaxies in cluster cores.
Among the star-forming galaxies, the SFE is higher at lower mass at any epoch, while the DS galaxies present both SFR and sSFR independent of mass.

In addition to the natural variations of SFE, the role of galactic outflows is also shown as contributing to the global evolution of the ZM relation,
even though its statistical effect is hard to assess on a global (rather than sub-grid) scale.
Thus, the effect of mechanisms helping to recycle metals more efficiently in the IGM, at
the same time hampering the star formation, cannot be neglected: processes in our simulations working towards this direction are  
either post-starburst strong winds triggering 
shocks, or dynamical heating due to the hierarchical growth, e.g. (dry) merging and galaxy encounters. 
The latter mechanisms have been shown in R08 to be rather relevant
above $z\simeq$1 (see BGG mass growth in their Fig. 9) and in dense environments and
this could actually explain the steeper slope profiles of cluster cores and
groups with respect to the cluster outskirts earlier than that epoch (Fig.\ref{sloDS}$b$).

In fact, massive DS systems in cluster outskirts show steep ZM at all epochs (Fig \ref{sloDS}$b$): 
such non-evolving ZM slope is a feature from our simulations for which we have not direct observational counterpart
as currently evidences generally come from mixed environments.
Moreover, they reach their maximum metallicity only at $z\sim$0, what suggests that here outflows acting to dissipate the
metal accumulation are less efficient than in cluster cores. 

These results indicate that during the dynamical phase of cluster pre-virialization some form of internal heating is relevant. 
Our simulations do not explicitely model AGN feedback: nevertheless, 
the role of AGN energy feedback in our simulations can be mimicked by the enhanced winds
originating from post-(gas-rich)-merger starbursts associated with simultaneous SN-II events.
Although these continuous winds are unrestricted in time by construction,
Antonuccio-Delogu \& Silk (2008) demonstrated that AGN feedback is able to suppress star formation in circumnuclear clouds during timescales 
much longer than that of the AGN cycle itself. 
Moreover, our super-wind scheme is likely as much effective as quasar-driven outflows in propagating metals at galactic scale, even though not
yet sufficient to quench innermost star formation by direct energy injection.
%
Then, the need for AGN feedback in reproducing the ZM relation would be less compelling than that demanded by the CM
relation, where the lack of a proper AGN modelling hinders correct age and SFR (hence colour) estimations at the centre of massive galaxies (see R08).

The combined effect of outflows of processed gas and internal heating would act thus as to shape the later stages of the DS building 
by shutting down star formation in more massive cluster galaxies, at the same time halting the metallicity uplift.
This evolution procedes starting from the inner galaxies following through the outer ones, with the latters not attaining yet
the abundance peak till present time; instead, massive galaxies in groups and cluster cores progressively saturate up to their maximum metal content, 
at around $z\simeq$0.5-0.2, while at the same time keep flattening their ZM relation (Fig.\ref{sloDS}$b$).
Conversely, in the low-mass regime, that begins to dominate the slope in groups and cluster cores since $z<$1,
the major feedback mechanisms acting on the ZM must rather be attributed to SNe, whose rate depends only and directly on the SFR 
at any epoch (see Scannapieco et al. 2006).

\bigskip

For what concerns the SF galaxies, 
they start migrating towards the DS as soon as a change in their SFE takes place, whether induced by shock-like events or by intrinsic nature;
there they will grow their metals following the rate of the DS systems, i.e. by simple ageing and mass loss of the stellar populations. 
At low $z$, the highest sSFR is concentrated at low masses (Fig.\ref{sSFR}),
where galactic winds shall be proportionally strong
enough to make metals evaporate without quenching the SF, until they also reach the 
point of SF shut off; after that, as it occurs in cluster cores,
they quickly increase their metal content, since there are no more mechanisms efficient in sweeping metals out to the
environment (see Fig. \ref{sloSF}$a$, third panel).

A scenario of enhanced star formation history at $z\lsim$1 in the low-to-intermediate mass bin is consistent with the epoch of last major mergers of our two 
clusters (between $z\sim$0.8 and 1.5), followed thence
by a steady increase of the mean metallicity in both core and outskirt cluster galaxies: such increase evidently lasts even
after the galaxy has entered the DS, until reaching a maximum at $z\sim$0.5.

\bigskip
In conclusion, we find that the overall (i.e. all-mass) SF galaxy population shows stronger evolution in both slope and scatter of the ZM
relation than DS galaxies, within the redshift range considered. In particular, the main SF sequence (as
emerging from Fig.\ref{SFR}) evolves in both slope and scatter as mainly driven by sSFR, that is SFE.
However, there are epochs like the post-merging phase undergone by our simulated clusters, when it is the combined effect of feedback and galaxy mergers 
that shapes its slope evolution by differentially affecting either mass regimes.

A possible emerging picture can then be proposed, where
wet galaxy mergers boost up total metallicity in more massive SF galaxies at $z\gsim$1 , 
while SN outflows deplete the less massive ones: both phenomena concur in raising the ZM relation slope at higher redshift;
at same epoch, dry mergers and/or metal-poor gas inflows regulate the metal balance in less active galaxies, resulting into a shallower ZM slope.
Conversely at $z\lsim$1, massive DS galaxies quench their residual star formation by internal shock heating, while low-mass ones keep ageing
passively: both stabilize at a slowed down rate of metal enrichment; SF galaxies instead have still gas available to feed star formation,
as likely result of metal-rich mergers mainly involving low-mass systems: this can explain the accelerated metal increase as well as the
relative flattening of their ZM slope, along with the positive correlation between metallicity and SFR.
We expect to further corroborate such interpretation by applying different feedback recipes in our simulations, as to better evaluate their effect
on the SFE.


\section*{Acknowledgments}
We wish to thank Antonio Pipino for fruitful discussions.
We also thank the anonymous referee for comments that helped improving the paper.

A.D.R. acknowledges support from 
FONDECYT - Proyecto de Iniciaci\'on a la Investigaci\'on No. 11090389
and from UNAB - Proyecto Regular No. DI-04-11/R.
I.G. acknowledges parcial support from FONDECYT through grant 11110501.
Finally we also thanks Patricia Ar\'evalo for ``The hour of writing".


\bigskip
\begin{thebibliography}{}

\bibitem[Andrews \& Martini (2013)]{am1} Andrews B.H. \& Martini P., 2013, ApJ 765, 140
\bibitem[Antonuccio-Delogu \& Silk (2008)]{van2} Antonuccio-Delogu V. \& Silk J., 2008, MNRAS 389, 1750
\bibitem[Arimoto \& Yoshii (1987)]{ay3} Arimoto N. \& Yoshii Y., 1987, A\&A 173, 23
\bibitem[Arrigoni et al. (2010)]{a4} Arrigoni M., Trager S.C., Somerville R.S., Gibson B.K., 2010, MNRAS 402, 173
\bibitem[Asplund et al. (2009)]{asp5} Asplund M., Grevesse N., Sauval A.J., Scott P., ARA\&A, 2009, 47, 481
\bibitem[Bell et al. (2004)]{bell6} Bell E.F., et al., 2004, ApJ 608, 752
\bibitem[Bower, Lucey \& Ellis (1992)]{ble7} Bower R.G., Lucey J.R. \& Ellis R.S., 1992, MNRAS 254, 601
\bibitem[Brooks et al. (2007)]{b8} Brooks A.M., et al., 2007, ApJL 655, 17
\bibitem[Calura et al. (2009)]{cal9} Calura F., Pipino A., Chiappini C., Matteucci F., Maiolino R., 2009, A\&A 504, 373
\bibitem[Calura \& Menci (2011)]{cal10} Calura F. \& Menci N., 2011, MNRAS 413L, 1
\bibitem[Daddi et al. (2007)]{da11} Daddi E., et al., 2007, ApJ 670, 156
\bibitem[Dav\'e, Finlator \& Oppenheimer (2011)]{dave12} Dav\'e R., Finlator K. \& Oppenheimer B., 2011, MNRAS 416, 1354
\bibitem[de la Rosa et al. (2011)]{dlr13} de la Rosa I.G., La Barbera F., Ferreras I., de Carvalho R.R., 2011, MNRAS 418L, 74
\bibitem[De Rossi, Tissera \& Scannapieco (2007)]{der14} De Rossi M.E., Tissera P.B. \& Scannapieco C., 2007, MNRAS 374, 323 
\bibitem[Elbaz et al. (2007)]{e15} Elbaz D., et al., 2007, A\&A 468, 33
\bibitem[Ellison et al. (2008)]{e16} Ellison S.L., Patton D.R., Simard L., Mc Connachie A.W., 2008, ApJ 672, L107	
\bibitem[Ellison et al. (2009)]{e17} Ellison S.L., Simard L., Cowan N.B., Baldry I.K., Patton D.R., McConnachie A.W., 2009, MNRAS 396, 1257
\bibitem[Erb et al. (2006)]{e18} Erb D.K., et al., 2006, ApJ 644, 813
\bibitem[Finlator \& Dav\'e (2008)]{f19} Finlator K. \& Dav\'e R., 2008, MNRAS 385, 2181
\bibitem[Foster et al. (2012)]{f20} Foster A.M., et al., 2012, A\&A 547, A79
\bibitem[Gallazzi et al. (2005)]{g21} Gallazzi A., Charlot S., Brinchmann J., White S.D.M., Tremonti C.A., 2005, MNRAS 362, 41
\bibitem[Gladders et al. (1998)]{g22} Gladders D.G., L\'opez Cruz O., Yee H.K.C., Kodama T., 1998, ApJ 501, 571
\bibitem[Hogg et al. (2004)]{h23} Hogg D.W., et al., 2004, ApJ 601L, 29
\bibitem[Kewley \& Ellison (2008)]{k24} Kewley L.J. \& Ellison S.L., 2008, ApJ 681, 1183	
\bibitem[Kitzbichler \& White (2007)]{k25} Kitzbichler M.G. \& White S.D.M., 2007, MNRAS 376, 2
\bibitem[Kobayashi, Springel \& White (2007)]{k26} Kobayashi C., Springel V. \& White S.D.M., 2007, MNRAS 376, 1465
\bibitem[Lamareille et al. (2009)]{l27} Lamareille F., et al., 2009, A\&A 495, 53
\bibitem[Lara-L\'opez et al. (2010)]{ll28} Lara-L\'opez M.A., et al., 2010, A\&A, 519A, 31L
\bibitem[Lilly et al. (1996)]{l29} Lilly S.J., Le Fevre O., Hammer F., Crampton, D., 1996, ApJL 460, 1
\bibitem[Lilly et al. (2013)]{l30} Lilly S.J., Carollo C.M., Pipino A., Renzini A., Peng Y., ApJ (2013)
\bibitem[Maiolino et al. (2008)]{m31} Maiolino R., et al., 2008, A\&A 488, 463
\bibitem[Mannucci et al. (2010)]{m32} Mannucci F., Cresci G., Maiolino R., Marconi A. \& Gnerucci A., 2010, MNRAS 408, 2115 
\bibitem[Matteucci (1994)]{m33} Matteucci F., 1994, A\&A 288, 57
\bibitem[McIntosh et al. (2005)]{m34} McIntosh D.H., et al., 2005, ApJ 632, 191
\bibitem[P\'erez-Montero et al. (2009)]{pm35} P\'erez-Montero E., et al., 2009, A\&A 495, 73
\bibitem[P\'erez-Montero et al. (2013)]{pm36} P\'erez-Montero E., et al., 2013, A\&A 549, A25 
\bibitem[Pipino \& Matteucci (2008)]{p37} Pipino A. \& Matteucci F., 2008, A\&A 486, 783
\bibitem[Pipino, Silk \& Matteucci (2009)]{p38} Pipino A., Silk J. \& Matteucci F., 2009, MNRAS 392, 475
\bibitem[Pipino et al. (2009)]{p39} Pipino A., Devriendt J.E.G., Thomas D., Silk J., Kaviraj S., 2009, A\&A 505, 1075
\bibitem[Romeo, Portinari \& Sommer-Larsen (2005)]{r40} Romeo A.D., Portinari L. \& Sommer-Larsen J., 2005, MNRAS 361, 983
\bibitem[Romeo et al. (2006)]{r41} Romeo A.D., Sommer-Larsen J., Portinari L., Antonuccio-Delogu V., 2006, MNRAS 371, 548
\bibitem[Romeo et al. (2008)]{r42} Romeo A.D., Napolitano N.R., Covone G., Sommer-Larsen J., Antonuccio-Delogu V. \& Capaccioli M., 2008, MNRAS 389, 13
\bibitem[Sakstein et al. (2011)]{s43} Sakstein J., Pipino A., Devriendt J.E.G., Maiolino R., 2011, MNRAS 410, 2203
\bibitem[Savaglio et al. (2005)]{s44} Savaglio S., et al., 2005, ApJ 635, 260	
\bibitem[Scannapieco et al. (2006)]{s45} Scannapieco C., Tissera P.B., White S.D.M., Springel V., 2006, MNRAS 371, 1125	
\bibitem[Sommariva et al. (2012)]{s46} Sommariva V., Mannucci F., Cresci G., Maiolino R., Marconi A., Nagao T., Baroni A., Grazian A., 2012, A\&A 539, 136
\bibitem[Sommer-Larsen, Romeo \& Portinari (2005)]{sl47} Sommer-Larsen J., Romeo A.D. \& Portinari L., 2005, MNRAS 357, 478
\bibitem[Thomas et al. (1999)]{t48} Thomas D., 1999, MNRAS 306, 655
\bibitem[Thomas et al. (2005)]{t49} Thomas D., et al., 2005, ApJ 621, 673
\bibitem[Thomas et al. (2007)]{t50} Thomas D., et al., 2007, IAUS 241, 546
\bibitem[Tissera, De Rossi \& Scannapieco (2005)]{t51} Tissera P.B., De Rossi M.E. \& Scannapieco C., 2005, MNRAS 364, L31
\bibitem[Tortora et al. (2011)]{t52} Tortora C., Romeo A.D., Napolitano N.R., Antonuccio-Delogu V., Meza A., Sommer-Larsen J., Capaccioli M., 2011, MNRAS 411, 627
\bibitem[Tremonti et al. (2004)]{t53} Tremonti C.A., Heckman T.M., Kauffmann G., Brinchmann J., Charlot S., White S., Seibert M., et al., 2004, ApJ 613, 898
\bibitem[Vale Asari et al. (2009)]{va54} Vale Asari N., Stasinska G., Cid Fernandes R., Gomes J.M., Schlickmann M., Mateus A., Schoenell W., 2009, MNRAS 396, L71
\bibitem[Yates, Kauffmann \& Guo (2012)]{y55} Yates R.M., Kauffmann G. \& Guo Qi, 2012, MNRAS 422, 215

\end{thebibliography}
\end{document}